\documentclass[a4paper,12pt]{article}
\usepackage{mathpazo}
\usepackage[top=4cm, bottom=4cm, left=3cm, right=3cm]{geometry}
\usepackage[utf8]{inputenc}
\usepackage{amsmath}
\usepackage[rgb,dvipsnames]{xcolor}
\usepackage{tikz}
\usetikzlibrary {arrows.meta,bending,positioning} 
\usepackage{standalone}
\usepackage{bm}
\usepackage{bbm}
\usepackage{graphicx}
\usepackage{amssymb}
\usepackage{amsthm}
\usepackage{mathrsfs}  
\usepackage{hyperref}
\usepackage[lofdepth,lotdepth]{subfig}
\usepackage[font=small,labelfont=bf]{caption}
\usepackage{authblk}
\usepackage{diagbox}
\usepackage{makecell}
\usepackage{adjustbox}

\usepackage{algorithm,setspace}
\usepackage[noend]{algpseudocode}
\algtext*{EndFor}
\algtext*{EndWhile}
\algtext*{EndIf}
\algtext*{EndFunction}

\algrenewcommand\textproc{}
\algnewcommand{\LeftComment}[1]{\Statex \(\triangleright\) #1}



\usepackage[style=ext-numeric,giveninits,doi=true,isbn=false,url=true,eprint=true,clearlang=true,sorting=nyt,maxbibnames=99,articlein=false,defernumbers=true,date=year]{biblatex}
\AtEveryBibitem{%
\ifentrytype{article}{
    \clearfield{urlyear}
    \clearfield{url}%
}{}
}
\renewbibmacro{in:}{%
  \ifentrytype{article}{}{\printtext{\bibstring{in}\intitlepunct}}}

\newcommand{\e}{\mathrm{e}}
\newcommand{\dd}{\mathrm{d}}

\DeclareMathOperator*{\argmax}{arg\,max}
\DeclareMathOperator*{\argmin}{arg\,min}

\newtheorem{theorem}{Theorem}
\theoremstyle{definition}
\newtheorem{video}{SM Video}

\def\scititle{
	Multicellular simulations with shape and volume constraints using optimal transport
}
\title{\bfseries \boldmath \scititle}
\author[1,2,$\dagger$]{Antoine \textsc{Diez}}
\author[3,$\dagger$]{Jean \textsc{Feydy}}
\date{}

\affil[1]{\small
RIKEN Center for Interdisciplinary Theoretical and Mathematical Sciences (iTHEMS)

RIKEN iTHEMS Wako Saitama 351-0198, Japan
\bigskip
}
\affil[2]{\small
Institute for the Advanced Study of Human Biology (ASHBi)

Kyoto University Institute for Advanced Study,

Kyoto University, 

Yoshida-Konoe-cho, Sakyo-ku, Kyoto 606-8501, Japan.
\bigskip
}
\affil[3]{\small
HeKA team, Inria Paris, Inserm, Université Paris-Cité,

PariSanté Campus, 2-10 rue d'Oradour-sur-Glane,

75015 Paris, France
\bigskip
}
\affil[$\dagger$]{
These authors contributed equally to this work.
}

\date{}

\begin{document} 

\maketitle

\begin{abstract} \bfseries \boldmath
Many living and physical systems such as cell aggregates, tissues or bacterial colonies behave as unconventional systems of particles that are strongly constrained by volume exclusion and shape interactions. Understanding how these constraints lead to macroscopic self-organized structures is a fundamental question in e.g. developmental biology. Here, we introduce a new framework to model particle systems with arbitrary volumes, dynamical shapes and deformability properties. Our method is grounded in optimal transport theory and its recent applications in incompressible fluid flows, crowd dynamics, and material sciences. Our approach supports a wide range of interaction and deformation mechanisms, while automatically taking care of the volume exclusion constraint with state-of-the-art numerical performance. We showcase the versatility of this approach through a series of experiments, demonstrating how it extends and refines results from previous approaches, with a special focus on challenging 3D situations in biophysics. Our Python code is freely available online. 
\end{abstract}

%

\section*{Introduction}
\noindent
From the smallest intracellular scale \cite{goodsell_living_1991} to the macroscopic population scale \cite{maury_handling_2011}, shape, deformations and congestion effects play a central role in the dynamics of living systems. For example, cells, bacteria and other microorganisms appear in a wide variety of shapes and may undergo large deformations, leading to complex spatial organization \cite{armstrong_cell_1989,farhadifar_influence_2007} and collective motion \cite{mehes_collective_2014}. Understanding and measuring these effects is a challenging but important experimental problem, especially in developmental biology \cite{ichikawa_ex_2022,ichbiah_embryo_2023}. Over the past decades, in complement to in vitro and in vivo experimental studies \cite{gonzalez-bermudez_advances_2019}, these questions have been studied using mathematical modeling. In silico experiments let us assess the impact of conjectured biophysical laws in minimal and cheap experimental settings where the shape and physical properties of each agent can be controlled individually. This poses a modeling and implementation challenge as that there is no unique nor simple mathematical way of representing dynamical shapes and multicellular systems. Depending on the objective, it requires to find an appropriate trade-off between the level of details to achieve sufficient physical accuracy, the numerical cost and the mathematical analytical potential. 

We propose a new mesoscale approach to simulate efficiently systems of tens to tens of thousands of soft-bodies. This roughly corresponds to the appropriate scale to study collective effects such as swarming or tissue organization during development. Before explaining the details, we briefly review other common models and refer to \cite{osborne_comparing_2017,brodland_computational_2004,vanliedekerke_simulating_2015} and to the introductions of \cite{mohammad_numerical_2022,kachalo_mechanical_2015} for an in-depth discussion of these approaches.

\begin{enumerate}
    \item \textbf{Point-particle systems} only consider the dynamics of spatial coordinates and finite-dimensional shape descriptors such as polarity vectors, aspect ratio etc. This formal simplicity is mathematically and numerically appealing but may not be fully realistic. They are typically more appropriate for very large systems at the statistical physics scale \cite{mehes_collective_2014}. 
    
    \item \textbf{Phase fields} models can represent arbitrary shapes as an indicator function which satisfies an Allen-Cahn equation \cite{nonomura_study_2012,nagayama_reaction_2023}. Strongly connected approaches consider the evolution of the boundary of a shape, defined as a {curve} \cite{saito_cell_2024,hiraiwa_dynamics_2011,ohta_dynamics_2017} or as the {level-set} of a potential function \cite{osher_level_2003,peskin_immersed_2002,yang_modeling_2008,mohammad_numerical_2022}. Contact interactions may be numerically and mathematically challenging and these approaches are usually most efficient for a small number of individuals. 
    
    \item \textbf{Cellular automata and lattice models} as introduced by Glazier and Graner \cite{glazier_simulation_1993,graner_simulation_1992}, consider a discretized spatial domain where each ``voxel'' represents either one biological cell or a portion of it. The voxel allocation and flipping dynamics are defined using custom energy minimization rules.

    \item \textbf{Voronoi tessellation} is an off-lattice tiling of the space by polygonal shapes which has been introduced in biology in \cite{honda_description_1978,sulsky_model_1984}. Although often realistic, recent experimental observations \cite{kaliman_limits_2016,miyazaki_mechanism_2023} have motivated the development of refined dynamical models \cite{bock_generalized_2010,bi_motilitydriven_2016,saye_voronoi_2011}. Our approach generalizes this line of work to flexible, non-polygonal shapes. 

    \item \textbf{Vertex models} define a tiling of the space via a set of vertices connected by straight edges in 2D or plane surfaces in 3D \cite{honda_threedimensional_2004}. The motion of each vertex typically results from an energy minimization hypothesis. More generally, defining an arbitrarily fine meshing of e.g. the cell membrane, the cell interior and/or the cell environment allows a particularly accurate physical description, but potentially at a higher numerical cost and mathematical complexity, in particular to treat topological changes. Such approaches are often referred as a {finite element} method \cite{chen_celllevel_2000,torres-sanchez_interacting_2022,zhao_dynamic_2017} or {Deformable Cell Models} (DCM) \cite{runser_simucell3d_2024,vetter_polyhoop_2024,madhikar_cellsim3d_2018}. 
\end{enumerate}

Importantly for the applications, most of these methods are available as open source softwares \cite{diez_sisyphe_2021,ghaffarizadeh_physicell_2018,kuang_morphosim_2023,starruss_morpheus_2014,swat_multiscale_2012,sego_general_2023,cooper_chaste_2020,torres-sanchez_interacting_2022,madhikar_cellsim3d_2018,vetter_polyhoop_2024,runser_simucell3d_2024}.
\smallskip

\textbf{Our approach.} Be it for cells in multicellular aggregates, microorganisms or pedestrians, the first measurable quantity is the volume. As a starting point, our method revolves around an independent volume constraint for each cell. Most of the approaches described above rather preserve cell volumes using soft constraints and relaxation forces towards a preferred size. Voronoi tessellation methods do not usually consider a volume constraint, or only provide little control on this quantity \cite{bock_generalized_2010}. As a first description, our method can be seen as a generalized Voronoi tessellation method with strict volume constraints. To achieve this goal, we rely on the notion of Laguerre tessellation which has recently appeared in various different contexts, in particular the simulation of incompressible fluid flows \cite{brenier_combinatorial_1989,gallouet_lagrangian_2018,levy_partial_2022}, of crowd motion \cite{maury_handling_2011,leclerc_lagrangian_2020} and the modeling of polycrystalline materials \cite{bourne_laguerre_2020,bourne_geometric_2023,buze_anisotropic_2024}. We also mention impressive applications in computer graphics \cite{degoes_power_2015,qu_power_2023,busaryev_animating_2012} where a particular case of Laguerre tessellations is known as power diagrams.

Although apparently fundamentally different, these situations are actually related to the theory of optimal transport. Originally developed for operations research and economics by Monge and Kantorovich, this theory describes the most effective way of allocating resources (canonically, piles of sand or flour) from one location to another while minimizing a transportation cost that is a function of the distance \cite{feydy_geometric_2020,peyre_computational_2019,santambrogio_optimal_2015}. The first connection between optimal transport and fluid mechanics is due to Brenier \cite{brenier_combinatorial_1989}. In the modern implementation of Brenier's ideas \cite{gallouet_lagrangian_2018,levy_partial_2022}, the notion of Laguerre tessellation appears as the natural spatial discretization procedure that preserves the core incompressibility constraint. Similar ideas are used in \cite{maury_handling_2011,leclerc_lagrangian_2020} to model crowd motion, where volume exclusion is crucial. More recently, and as a direct inspiration for our work, Laguerre tessellations and optimal transport have been applied to materials science \cite{alpers_generalized_2015,bourne_laguerre_2020,bourne_geometric_2023,buze_anisotropic_2024}, where, similarly to biological cell aggregates, some materials like steel are made up of a collection of tiny ``crystals'' with various shapes and sizes. On the implementation side, we leverage the divide-and-conquer ``multiscale'' strategy \cite{feydy_geometric_2020,schmitzer_stabilized_2019,merigot_multiscale_2011} commonly adopted in optimal transport. Our implementation relies on massively parallel Graphics Processing Units (GPUs), with a state-of-the-art solver accessible through a convenient Python interface \cite{charlier_kernel_2021}.

Although our model is formally a tessellation model, we show that it shares important properties with point-particle systems, level-set methods and vertex models, which thus also suggests a novel optimal transport point of view for these methods. As a consequence, our model is remarkably versatile: within the same framework, we can represent individual particles with arbitrary shapes (as in level-set methods), their collective motion (as in point-particle systems) and much denser tissue-like aggregates whose dynamics is ruled by surface tension and other contact-based interactions (typically treated with DCMs). The graphical abstract Fig.~\ref{fig:graphical_abstract} summarizes our approach.

\section*{Results}

\subsection*{Static model: Laguerre tessellation}

We consider a set of $i=1,\ldots,N$ particles, each of them defined by the couple $({x}_i,v_i)$ of its position, in a given domain $\Omega$, and its volume. By convention, we assume that the total volume of the domain is normalized to 1. The total volume occupied by the particles should thus be $V = v_1+\ldots + v_N \leq 1$. The main idea of the present article is to model the space occupied by each particle $i$ as a \textit{Laguerre cell} $\mathscr{L}_i$, defined as the set of points which satisfy the following set of inequalities: 
\begin{equation}\label{eq:Laguerre}\mathscr{L}_i = \{{x}\in \Omega,\,\,c({x},{x}_i) - w_i \leq c({x},{x}_j) - w_j\,\,\text{for all}\,\, j\}.\end{equation}
\begin{enumerate}
    \item The function $c : \Omega\times\Omega\to [0,+\infty)$ is an arbitrary function, called \textit{cost function}, which will encode both the shape and the deformability properties of each individual particle. One can instead consider $N$ functions $c_i : \Omega \to [0,+\infty)$ but we will mostly consider the case $c_i({x}) = c({x},{x}_i)$ here. For the time being, a typical example to keep in mind is the $L^2$ cost defined as the square of the distance function: $c({x},{y}) = |{x}-{y}|^2$. 
    \item The \textit{Kantorovich potentials} $w_1,\ldots,w_N$ are uniquely defined in order to satisfy the volume constraint $|\mathscr{L}_i| = v_i$.
\end{enumerate}

For a large class of cost functions and any positions ${x}_i$, the Laguerre tessellation \eqref{eq:Laguerre} can be shown to be the unique solution of the following constrained minimization problem on the set of partitions of $\Omega$: 
\begin{equation}\label{eq:totalcost}\mathcal{T}_c = \min_{(\mathscr{L}_i)_{i=1,\ldots,N}}\left\{ \sum_{i=1}^N \int_{\mathscr{L}_i} c({x},{x}_i)\dd {x},\,\,\,\text{with constraints}\,\,\,|\mathscr{L}_i| = v_i\right\}.\end{equation}
In optimal transport theory, such partition is then understood as an assignment problem, where each point ${x}\in\Omega$ is assigned to one of the ${x}_i$ at a cost $c({x},{x}_i)$ (see Supplementary for more details). Using this interpretation, one of our main contributions is a fast numerical method to compute the Kantorovich potentials~$w_i$. 

When $c$ is the $L^2$ cost and $w_i=0$, we recover the standard definition of a Voronoi diagram. Our Laguerre model can thus be understood as a generalized Voronoi tessellation with strict volume constraints. Generalizing the distance function into an arbitrary function $c$ is a key modeling idea which allows arbitrary boundary shapes between neighboring particles, since they are defined by algebraic equations of the form 
\[\mathscr{L}_i\cap\mathscr{L}_j\subset \{c({x},{x}_i) - w_i = c({x},{x}_j) - w_j\}.\]
With the $L^2$ cost, we recover polygonal shapes, but other choices may lead to more curved boundaries typically observed in biology (Fig.~\ref{fig:shapes}A). In the case $V<1$, we model the empty space as one additional particle ${x}_0$ (at an arbitrary location) with volume $1 - V$ and associated to the zero cost $c_0({x})~=~0$. The shape (or boundary) of an isolated particle thus corresponds to a level set of the cost function:
\[\partial\mathscr{L}_i = \{{x}\in\Omega,\,\, c({x},{x}_i) = w_i - w_0\}.\]
The $L^2$ cost corresponds to spherical shapes but any shape can actually be realized as the level set of a custom function $c$ (Fig.~\ref{fig:shapes}B,C). We will show later that the cost function also encodes how easily these shapes can deform due to collisions in a dynamical framework. This common framework thus connects the so-called generalized Voronoi tessellation models and level-set methods. Most importantly, the optimal transport theory provides a natural way to treat volume constraints, which are otherwise enforced in a soft manner or only heuristically.

\subsection*{Dynamical model}

In a dynamical framework, we first consider the following first-order gradient descent equation for the particles' locations:
\begin{equation}\label{eq:incompressibilityforce}\dot{{x}}_i = - \tau_i\nabla_{{x}_i} \mathcal{T}_c = -\tau_i \int_{\mathscr{L}_i} \nabla_{{x}_i} c({x},{x}_i) \dd{x}, \end{equation}
where $\tau_i\geq0$ is an arbitrary gradient step and the last equality is proved in \cite{xin_centroidal_2016}. In fluid mechanics \cite{gallouet_lagrangian_2018,levy_partial_2022,brenier_combinatorial_1989}, this is the analog of an \textit{incompressibility force} exerted on each microscopic fluid element. In the present context, this motion leads to repulsion interactions between the particles (SM Video~\ref{video:incompressibility}) since the absolute minimum of the total cost $\mathcal{T}_c$ \eqref{eq:totalcost} is achieved when the particles are all isolated (if space allows). For the $L^2$ cost, the incompressibility force points in the direction of the barycenter of the Laguerre cell. Unlike point-particle models, this force is not a sum of binary forces between two locations: it takes into account the actual shapes and contact surfaces between all the particles.

If the cost function only depends on the connecting vector (which will always be the case), i.e. $c(x,x_i)\equiv c(x_i - x)$, then $\nabla_{x_i} c(x,x_i) = -\nabla_{x} c(x,x_i)$ and Stokes' theorem gives an alternative expression for the incompressibility force as an integral over the boundary $\partial\mathscr{L}_i$ of $\mathscr{L}_i$: 
\[- \nabla_{{x}_i} \mathcal{T}_c = -\int_{\partial\mathscr{L}_i} c(x,x_i)\vec{n}\dd \sigma(x),\]
where $\vec{n}$ denotes the inward normal and $\sigma$ the surface measure. This expression shows that the incompressibility force can also be interpreted physically as a resulting internal pressure force, where the pressure on each surface element is proportional to the cost function. Large deformations are thus naturally penalized when the cost is a non decreasing function of the distance. It also implies that the equations of motion \eqref{eq:incompressibilityforce} preserve the total momentum in the absence of active forces, boundary and when $\tau_i\equiv1$. Indeed, summing over $i$ and denoting by $\Gamma_{ij} = \mathscr{L}_i \cap \mathscr{L}_j$ the boundary between two Laguerre cells, 
\[\sum_{i=1}^N \dot{x}_i = \sum_{\{i,j\}} \int_{\Gamma_{ij}} \{c(x,x_j) - c(x,x_i)\}\,\vec{n}\,\dd\sigma,\]
where the sum is over all unordered pairs of indices and $\vec{n}$ points towards the $i$-th cell. By definition of the Laguerre cells, the integrand is constant on $\Gamma_{ij}$, equal to $w_j - w_i$ so that
\[\sum_{i=1}^N \dot{x}_i = \sum_{i=1}^N w_i \int_{\partial\mathscr{L}_i} \vec{n} \dd\sigma = 0,\]
since $\partial\mathscr{L}_i$ is a closed surface. 

We can take advantage of this clear mathematical structure to incorporate any other force or noise terms. We will consider general first-order (stochastic) differential equations systems of the form
\begin{equation}\label{eq:generalequation}\dot{{x}}_i = - \tau_i\nabla_{{x}_i} \mathcal{T}_c  + {F}^{\mathrm{point}}_i({x}_1,\ldots,{x}_N) + {F}_i^{\mathrm{surf}}(\mathscr{L}_1,\ldots,\mathscr{L}_N) + \dot{{\xi}}_i.\end{equation}
The force term ${F}^{\mathrm{point}}_i$ includes all the external or interaction forces which depend only on the locations, as in point-particle models: e.g. gravity, self-propulsion, attraction etc. The force term ${F}^{\mathrm{surf}}_i$ on the contrary may depend on the shapes of the particles and can typically model pressure forces and surface tension effects along each interface. Stochastic effects can be included in the force terms or by adding a stochastic noise ${\xi}_i$. Note that for any dynamics, the non-overlapping and volume constraints are automatically ensured at all times by the definition of the Laguerre cells \eqref{eq:Laguerre}.

One can also consider dynamical volumes $v_i$ or cost functions (i.e. shapes) $c_i$. Just as for the position ${x}_i$, the time derivative of the volume $\dot{v}_i$ may depend on all the components of the model thus allowing arbitrary growth behaviors. For the dynamics of the cost we introduce two (non-exclusive) main modelling ideas. 
\begin{enumerate}
    \item \textit{Parametrized costs:} $c_i({x}) = c({x},{x}_i;\mathcal{P}_i)$ where $\mathcal{P}_i$ is an arbitrary set of parameters: typically they will model an orientation. A prototypical example is 
    \begin{equation}\label{eq:ellipsoidcost}c({x},{x}_i;\Sigma_i) = ({x} - {x}_i)^{\mathrm{T}}\Sigma_i^{-1}({x}-{x}_i),\end{equation}
    where $\Sigma_i$ is a covariance matrix, which in this case imposes an ellipsoid shape. The parameters $\mathcal{P}_i$ can have an arbitrary dynamics, as in classical active particle models, but can also possibly depend on the shapes $\mathscr{L}_i$.
    \item \textit{Level-set potential functions:} $c_i = \e^{-\varphi_i}$ where $\varphi_i = \varphi_i^0 + f_i$ is a potential functions which is a modification of a base potential $\varphi_i^0$ with a perturbation $f_i$. While $\varphi^0_i$ defines the base shape of the particle, the perturbation $f_i$ introduces a deformation bias in arbitrary (potentially time-varying) directions, typically given by external clues such as a chemotactic field.
\end{enumerate}

Note that the optimization step ensures the strict volume constraint thanks to the unique choice of the potentials $w_i$ in \eqref{eq:Laguerre}, while the incompressibility force \eqref{eq:incompressibilityforce} tends to restore the preferred shapes. The approach thus differs from Voronoi-based models \cite{bi_motilitydriven_2016} where the potentials $w_i=0$ are always kept constant and the incompressibility constraint is instead enforced using an additional energy minimization principle. However, in both models, the main parameters are the centroid positions $x_i$ that follow equations of motion of the form \eqref{eq:generalequation}. Finally, let us mention that since the cell shapes are instantaneously computed as a Laguerre cell (i.e. the relaxation of cell shapes occurs within a single simulation step), the framework implicitly assumes that the relaxation of the interface shape is much faster than the relaxation of the cell-center positions $x_i$. Under this approximation, this model does not describe the intrinsic relaxation dynamics of the interface shape itself.

In the following, we provide several examples to showcase the applicability and versatility of our model in comparison to other computational models. All these examples are based on \eqref{eq:generalequation}, The detailed numerical values of the parameters as well as implementation details regarding the computation of the cost, of the interaction forces and of the boundary conditions are gathered in the Methods section.  

\subsection*{Soft-body simulations: first examples and benchmark}

As a first illustration, we start with the $L^2$ cost and consider a simple free-fall motion corresponding to ${F}_i^\mathrm{point} = -{e}_z$ and ${F}_i^\mathrm{surf}=0$. To emphasize the deformations, we consider a hourglass domain $\Omega$. In this active Voronoi model, particles tend to keep spherical shapes and their contact surfaces are planes (Fig.~\ref{fig:benchmark}A).

As a classical test case in computational biology \cite{nonomura_study_2012,farhadifar_influence_2007,barton_active_2017,kachalo_mechanical_2015,runser_simucell3d_2024}, we consider the growth of a cellular aggregate, with no external forces and only the volume exclusion repulsion force modeled by \eqref{eq:incompressibilityforce}. Initially a single cell grows linearly. When it reaches a target volume and after a random exponential time, this cell divides along a random division plane in two daughter cells with equal half volumes and which follow the same dynamics. The volume exclusion interactions lead to an exponentially growing spherical aggregate (Fig.~\ref{fig:benchmark}B). We end the simulation when its volume entirely fills a given box domain, reaching $N=50,000$ particles. Using our implementation, the total computation time is about one day.

As a more detailed benchmarking experiment, we consider a system of self-propelled 3D deformable ellipsoids constrained in a box domain (Fig.~\ref{fig:benchmark}C). The orientation of each particle is given by its covariance matrix $\Sigma_i$ which defines both its direction of motion and aspect ratio. The ellipsoidal shape is enforced by the parametrized cost \eqref{eq:ellipsoidcost}. We consider a strongly packed situation: the covariance matrix of each particle is defined at each time using the Principal Component Analysis (PCA) of its current Laguerre cell. This results in a run-and-tumble motion (SM Video~\ref{video:benchmark}). The Table \ref{tab:benchmark} shows the simulation time for 2000 iterations depending on the number of particles $N$ and the space discretization grid~$M$ (see Methods). The complexity is linear in $N$ and $M^d$ but the GPU implementation is sub-optimal for small $N$.

\subsection*{Emergence of orientational order for rod-shape active particles}

Oriented particles with anisotropic shapes can be realized in our optimal transport framework by considering appropriate cost functions. In dimension 2, any shape defined by a polar equation $r=r_0(\theta)$ can be encoded by the cost 
\begin{equation}\label{eq:costshape} c({x},{x}_i) = \left(\frac{|{x}-{x}_i|}{r_0(\theta({x},{x}_i))}\right)^\alpha,\end{equation}
where $\theta({x},{x}_i)$ denotes the polar angle of the vector ${x}-{x}_i$ and $\alpha>0$ is a hardness parameter whose influence will be illustrated later. The extension to dimension~3 is straightforward. An important example is the spherocylinder shape of 2D bacilli and rod-shape polymers. Given a long axis vector $(\cos\theta_0,\sin\theta_0)^\mathrm{T}$ and an aspect ratio $s=a/b\geq1$, this shape can be realized using the cost associated to the function
\begin{equation}\label{eq:spherocylindercost}
r_0(\theta) =\left\{
\begin{array}{l}
\frac{b}{|\sin(\theta-\theta_0)|},\quad\text{if}\,\,|\tan(\theta - \theta_0)| \geq \frac{1}{s-1},\\
b\left(|\cos(\theta - \theta_0)|(s-1) + \sqrt{1 - ((s-1)\sin(\theta-\theta_0))^2}\right),\quad \text{otherwise.}
\end{array}
\right.
\end{equation}
Using \eqref{eq:costshape}, we model a system of self-propelled swarming bacteria. The cost is parametrized by the long-axis vector which also defines the direction of motion. It is defined at each time as the leading vector of the PCA of the current Laguerre tessellation. This motion corresponds to ``bending'' and turning effects due to the collisions. With this simple rule, the emergence of orientational order (or alignment) is observed from the sole volume exclusion interactions, as classically hypothesized in polymer physics (Fig.~\ref{fig:shapes}B with $N=300$ and Fig.~\ref{fig:nematicorder}C with $N=150$). A similar idea has recently been developed in \cite{leech_derivation_2024} in a computational model of deformable ellipsoidal particles representing fibroblasts.

This emergent order can classically be evaluated quantitatively by computing the time evolution of the nematic order parameter $\phi = \frac{d}{d-1}\lambda\in[0,1]$ where $\lambda$ is the largest eigenvalue of the matrix obtained by averaging the individual Q-tensor matrices $Q_i = n_i\otimes n_i - \frac{1}{d}\mathrm{Id}$, with the dimension $d=2$ and $n_i$ the long-axis vector of particle $i$. For the larger system (Fig.~\ref{fig:shapes}B), a sharp transition is observed around time $t=12$ and orientational order is preserved until $t=25$ (Fig.~\ref{fig:nematicorder}A). However, for this set of parameters, we have observed that the orientational order is typically not persistent over extended intervals of time but rather fluctuates following the the successive formation and disappearance of ordered clusters (SM Video~\ref{video:rodshape}). A persistent orientational order can be observed in the smaller ($N=150$) but denser system (see Fig.~\ref{fig:nematicorder}B,C for a simulation of such system until $t=600$).

\subsection*{Fluid-solid phase transition}

In this section we consider soft spherical particles of equal size associated to power costs of the form 
\begin{equation}\label{eq:powercost}c_\alpha({x},{y}) = \lambda_\alpha^{-1} |{x}-{y}|^\alpha,\end{equation}
where $\lambda_\alpha>0 $ is a normalizing factor (see Methods). Here the exponent $\alpha$ plays the role of a deformability parameter. Indeed when $\alpha>0$ is large, the cost $c({x},{x}_i)$ of assigning a point ${x}$ to the cell $i$ is close to zero for points near ${x}_i$, but it increases sharply for points farther away. On the contrary, when $\alpha\to0$, the cost function becomes flatter and thus assigns a similar cost to any point. Hence a small $\alpha$ allows large deformations while a larger $\alpha$ penalizes deformations.

As in \cite{bi_motilitydriven_2016,saito_cell_2024}, we confirm that the cell deformability parameter $\alpha$ drives a fluid-solid phase transition in a 2D system of active Brownian particles. We simulate the following system with $N=250$ in the square periodic domain $\Omega=[0,1]^2$:
\begin{equation}\label{eq:abp}\dot{{x}_i} = c_0 {n}_i - \tau\nabla_{{x}_i}\mathcal{T}_{c_\alpha},\quad \dot{{n}}_i = \sqrt{2D} \dot{{\xi}}_i,\quad |{n}_i|=1,\end{equation}
where $c_0>0$ is the self-propulsion speed and $D$ is the diffusion of the direction of motion.
A single point particle $(\tau=0)$ has a purely diffusive behavior with a Mean-Square Displacement (MSD) growing linearly in time at speed $D_{\text{thr}} = 2c_0^2/D$. We compute the effective diffusion coefficient $D_{\text{eff}}$ of the shape constrained system $(\tau>0)$ as the slope of the best linear curve fitting the MSD of the centroids ${x}_i$ over time (in the least squares sense, Fig.~\ref{fig:abp}C) after unwrapping the periodic trajectories to the full domain. A few initial steps of Lloyd algorithm ensure that the particles are initially well-scattered. To avoid boundary effects, we simulate the system until the time $0.5/c_0$. A longer simulation time produces similar results (Fig.~\ref{fig:msdloglog}A). The phase diagram (Fig.~\ref{fig:abp}A) shows the normalized value $\overline{D}_{\text{eff}} = D_\text{eff}/D_\text{thr}$ for small P\'{e}clet numbers and constant values $D=20$ and $\tau=3$.

The diffusive behavior can be evaluated by the coefficient of determination $r^2$ of the linear fit (Fig.~\ref{fig:abp}C). The deformation of a Laguerre cell $\mathscr{L}_i$ is measured as in \cite{bi_motilitydriven_2016} by the shape index $\sigma_i=\frac{|\partial \mathscr{L}_i|}{2\sqrt{\pi}|\mathscr{L}_i|}\geq1$ with $\sigma_i=1$ corresponding to a disk (Fig.~\ref{fig:abp}C). 

In the fluid phase, the system keeps a diffusive behavior ($r^2>0.9$) despite the volume exclusion interactions but with a lower diffusion coefficient $\overline{D}_\text{eff}~<~1$. Note that for $c_0$ large and $\alpha$ small, the shape constraint is very weak which produces unrealistic non-connected or engulfed shapes while $\overline{D}_\text{eff}$ converges to 1 (Fig.~\ref{fig:abp}B). In the solid phase the effective diffusion vanishes and the system reaches a stable hexagonal hard-sphere packing final configuration with an average shape index $\langle\sigma_i\rangle$ close to 1. The transition between the fluid and the solid phases (Fig.~\ref{fig:abp}A white plain line) is computed from the shape index diagram (Fig.~\ref{fig:abp}C) as the level set $\sigma_i = 1.015$ where the Laguerre cells become indistinguishable from perfect disks at the resolution considered. In the solid phase, the MSD saturates and the $r^2$ value of the linear fit drops (Fig~\ref{fig:abp}C). In addition, a log-log plot of the MSD that shows that the anomalous diffusion exponent remains close to 1 for most values of $\alpha$ is shown in Fig.~\ref{fig:msdloglog}B.

The dashed white line corresponds to the level-set of an average shape index $\langle \sigma\rangle = \frac{3.81}{2\sqrt{\pi}}$ which is advocated in \cite{bi_motilitydriven_2016} as a transition point between a fluid and a soft-fluid phase. In our model, as in the contour model \cite{saito_cell_2024}, this valued does not seem to correspond to a sharp transition of the effective diffusion. In \cite{saito_cell_2024} this special value is rather associated to the percolation of topological defects but as it is not the main goal of the present article, we leave this point for later work.  

Such model can be used to simulate crowds and jamming phenomena due to volume exclusion as well as sorting between heterogeneous populations of more or less soft and heavy particles. Such simulations are presented in the Supplementary Material. Moreover, unlike other computational approaches, our model is particularly well-suited for coarse-grained mathematical analysis. Based on theoretical results in optimal transport theory, a formal derivation presented in the Supplementary Text leads to a mean-field approximation of the particle system \eqref{eq:abp}. It takes the form of a system of Partial Differential Equations that describes the evolution of the density of particles $f(x,n)$ at a position $x$ and with the orientation $n$
\begin{subequations}
\begin{align}
    &\partial_t f(x,n) + c_0\nabla_x f = \tau \nabla_x \cdot \left( \nabla_x\Phi f\right)  + D\Delta_n f, \\ 
    &\det\big(\mathrm{I} - \nabla^2_x\ell^*(\nabla_x\Phi)\nabla_x^2\Phi\big) = f.
\end{align}
\end{subequations}
The potential $\Phi(x)$ is the solution of a nonlinear Monge-Amp\`ere equation, which is known in other contexts in physics \cite{dephilippis_monge_2014,brenier_geometric_2004}. The function $\ell^*$ is the Legendre transform of the function $\ell(x) = |x|^\alpha$.

\subsection*{Deformation-driven motion}

In the level-set formalism \cite{yang_modeling_2008} the boundary of each individual particle is defined as the level-set of a potential function $\varphi$, usually initially chosen as the signed distance from the boundary. In our framework, this representation is equivalent to taking a cost of the form $c = \e^{-\varphi}$ (since the cost is an intrinsically nonnegative multiplicative quantity). Level-set methods are particularly adapted to model the motion of biological cells induced by the deformation of their membrane, driven for instance by an external chemotactic field. Here, as in \cite{yang_modeling_2008}, our goal is not a full description of how this field drives the internal reorganization of the cell's actin cortex but we rather aim at a minimal phenomenological model able to reproduce different scenarios of cell motion.

We consider a chemo-attractant density $u({x})$ in $\Omega$ and we classically assume that a particle located at ${x}_i$ can sense the local gradient $\delta u ({x},{x}_i)$ along the directions ${x}-{x}_i$, defined through a finite difference formula. To model the deformations induced by this field, the dynamics of $\varphi$ could typically be defined by an advection equation modeling the forces acting on the boundary. However, here we consider a simpler approach where a deformation term is added to the potential defining a spherical $L^2$ Laguerre cell:
\begin{equation}\label{eq:costpotentialchemotaxis}\delta u ({x},{x}_i) = \frac{u({x})-u({x}_i)}{|{x}-{x}_i|},\quad \varphi({x},{x}_i) = -2\log |{x} - {x}_i| + \beta f\big(\delta u({x},{x}_i)\big),\end{equation}
for some constant $\beta>0$ and some function $f$ which models how the gradient affects the deformation. This deformation can be orthogonal to or along the chemo-attractant gradient \cite{yang_modeling_2008}. Denoting by $\delta_+ = \max(0,\delta)$ and $\delta_- = \max(0,-\delta)$ the positive and negative parts of a number $\delta$, two choices for $f$ are thus 
\begin{equation}\label{eq:bias}
f(\delta) = -\delta_+,\quad f(\delta) = \delta_-^2 + \delta_+ \end{equation}
The first $f$ lowers the potential in the directions of large nonnegative gradients. This produces elongated shapes (Fig~\ref{fig:shapes}C and SM Video~\ref{video:chemo_long}).  The second $f$ increases the cost in the direction of both positive and negative gradients but more in the latter case. This produces fan-like shapes (Fig~\ref{fig:shapes}C and SM Video~\ref{video:chemo_fan}). In both situations, we consider the simplest possible dynamics given by \eqref{eq:incompressibilityforce}. Since the morphological changes are biased in the direction of increasing gradient, this choice induces chemotactic migration, with the two different cell shape deformations. Both situations are comparable to the ones obtained with the level-set method in \cite{yang_modeling_2008} and are in agreement with the experimental observations therein. 

An important point to note is that not only the volume is preserved regardless of the deformation, but our model also includes non-overlapping and cell-cell repulsion. The morphological changes resulting from these mechanical interactions would typically be difficult to model in the level-set formalism, where mostly a single cell or non-interacting cells are considered. They are inherently included in our modeling framework.

\subsection*{Cell sorting via surface interactions}

So far, Laguerre cells have been used to model soft bodies with internal pressure forces that maintain a preferred shape, with a response to deformations that can typically be controlled using the hardness parameter $\alpha$. In some contexts such as biological cells and bubbles, it is important to also model adhesion phenomena that typically depend on surface tension effects. To do so, another set of forces and parameters need to be introduced. In this section, we thus model large aggregates of deformable spheres using Laguerre cells defined by a simple weighted $L^2$ cost but interacting via the following forces computed along each interface: 
\begin{equation}\label{eq:interfaceforce}
    c({x},{x}_i) = \frac{\gamma_{i0}}{R_i} |{x} - {x}_i|^2,\quad {{F}}_{i\leftarrow j} = \int_{\Gamma_{ij}} \left(\gamma_{ij} |\kappa| + \frac{\eta_{ij}}{|{x}_i - {x}_j|}\right){\vec{n}}\, \dd \sigma.
\end{equation}

The force ${{F}}_{i\leftarrow j}$ results from elementary pressure-like interactions depending on the local mean curvature $\kappa$ along each interface $\Gamma_{ij} = \mathscr{L}_i\cap \mathscr{L}_j$ between the Laguerre cells $i$  and $j$ (with possibly $j=0$). The vector ${\vec{n}}$ denotes the inward normal of $\mathscr{L}_i$ so this force is always a repulsion force. By moving the centroid~${x}_i$ away from its interface, the first term reduces the local curvature while the second reduces the interface area. The parameters $\gamma_{ij},\eta_{ij}>0$ have the dimension of surface tensions. When $j=0$ (interaction with the medium), we set $\eta_{i0} = 0$ so that only the curvature term remains. The surface tension $\gamma_{i0}$ between a cell and the medium is also assumed to control the relative softness between the particles by defining the weight of the cost function. The cost is non-dimensionalized by dividing by the radius $R_i$ of the particle $i$. Larger particles with a smaller surface tension parameter thus appear relatively softer. The choice of the $L^2$ cost ensures that each interface has a constant curvature. The total force exerted on ${x}_i$ is finally defined as 
\[{{F}}^\text{surf}_i = \sum_{j=0}^N {{F}}_{i\leftarrow j} - \frac{1}{N_i} \sum_{k\in\mathcal{N}_i} {{F}}_{k\leftarrow 0},\]
where $\mathcal{N}_i$ is the connected aggregate to which $i$ belongs and $N_i$ denotes the number of cells in $\mathcal{N}_i$. The second term on the right-hand side is a correction term. Indeed, although for two Laguerre cells, ${{F}}_{i\leftarrow j} = -{{F}}_{j\leftarrow i}$, there might be nonzero momentum gain coming from the interaction with the medium so we chose to evenly split it within the cell aggregate to restore the conservation of momentum. Note that this is a consequence of the simplifying approach to treat the medium as one single particle with zero cost. Alternatively, the medium could be treated as an incompressible fluid discretized using a Laguerre tessallation as in \cite{gallouet_lagrangian_2018,levy_partial_2022} which would preserve momentum but would be numerically much more costly.

\begin{figure}
    \centering
    \includegraphics[width=0.98\linewidth]{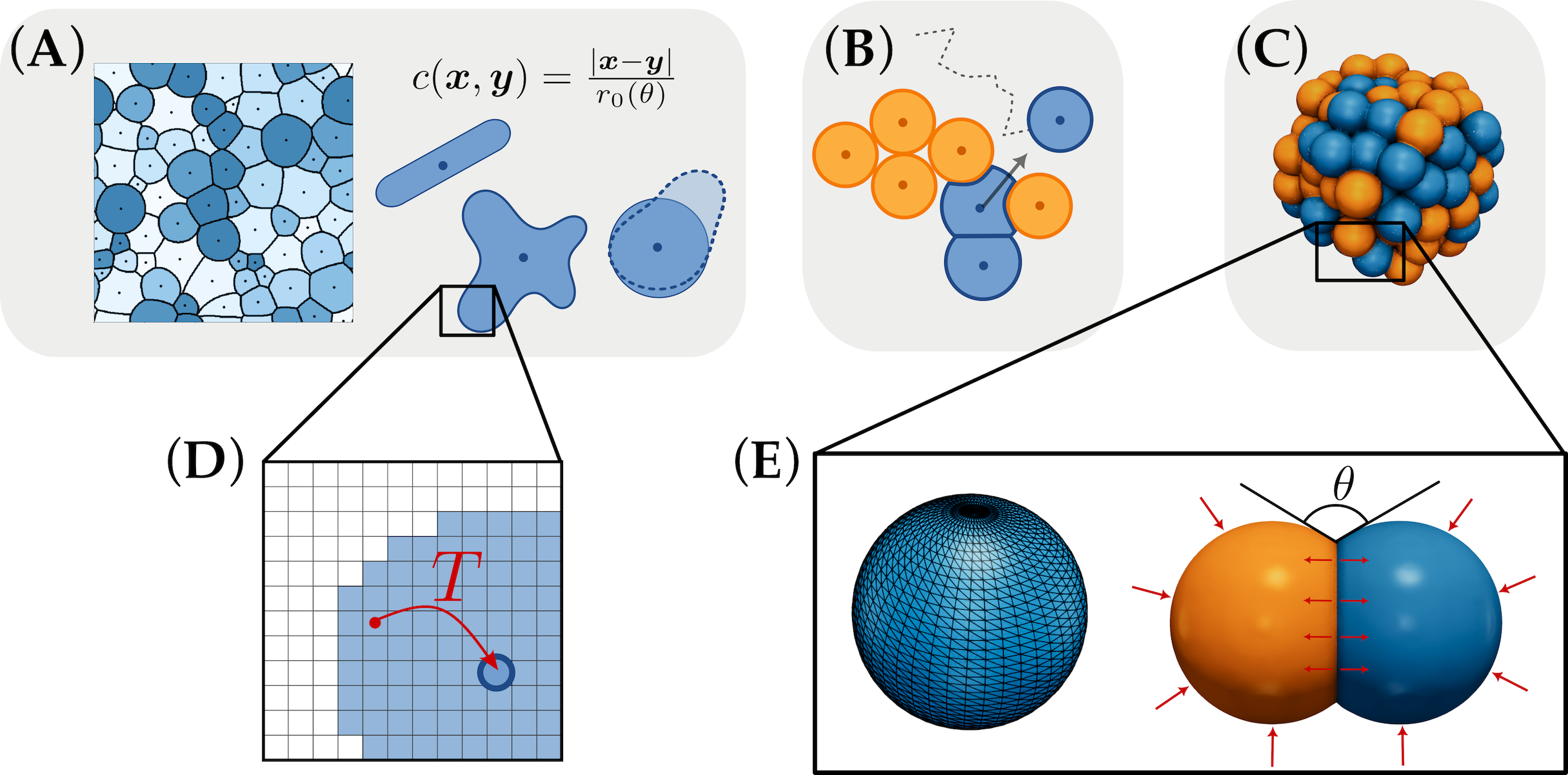}
    \caption{\textbf{Graphical abstract.} \textbf{(A)}~Laguerre tesselations generalize Voronoi diagrams and level-set approaches with volume, shape and deformation constraints encoded in a cost function $c$ which can be customized and dynamic. \textbf{(B)}~Any active point-particle model can be implemented with additional arbitrary softness and deformation properties. \textbf{(C)}~The framework is independent of the dimension and is implemented in 3D. \textbf{(D)}~Laguerre tessellations are computed as the solution of a semi-discrete optimal transport problem on a discrete grid, resulting in a map $T$ which assigns each voxel to a cell. \textbf{(E)}~In 3D, a meshing of each cell boundary is computed in order to implement surface tension effects and cell sorting mechanisms.}
    \label{fig:graphical_abstract}
\end{figure}

This model satisfies the Young-Dupr\'e relationship which states that the equilibrium configuration of two identical particles is a cell doublet made of two spherical caps (Fig.~\ref{fig:graphical_abstract}E) intersecting with a contact angle $\theta\in[0,\pi]$ defined by
\[\cos \frac{\theta}{2} = \frac{\eta}{2\gamma}.\]
We further validate this model by reproducing various cell sorting phenomena observed in biology which are commonly used as a standardized test case for computational models \cite{brodland_differential_2002,brodland_computational_2004,mohammad_numerical_2022,graner_simulation_1992,glazier_simulation_1993,belousov_when_2023,sulsky_model_1984} since the work of Chen and Brodland on the so-called Differential Interfacial Tension Hypothesis (DITH) \cite{chen_celllevel_2000,brodland_mechanics_2000a,brodland_differential_2002}. While most models in the literature are intrinsically 2D, we showcase the applicability of our approach in a fully 3D setting.

Following Chen and Brodland, we thus consider an initial homogeneously mixed aggregate with two types of cells denoted by $b$ and $o$ (respectively for blue and orange), embedded in a medium (Fig.~\ref{fig:graphical_abstract}C). The parameters $\gamma$ and $\eta$ are assumed to depend only on the cell type. Our model thus has six parameters denoted by $\gamma_{oo},\gamma_{bb},\gamma_{ob},\eta_{oo},\eta_{bb},\eta_{ob}$ plus two surface tension parameters denoted by $\gamma_b$ and $\gamma_o$ between each type and the medium. Since the interface between two cells of the same type has zero curvature, we can set $\gamma_{bb}=\gamma_{oo}=0$. To investigate the phase diagram of this model, we first define several dimensionless parameters. First, the compaction parameters respectively for orange and blue aggregates and a virtual orange aggregate in a medium similar to blue particles are defined using Young-Dupré relation as:
\[k_o = \frac{\eta_{oo}}{2\gamma_{o}}, \quad k_b = \frac{\eta_{bb}}{2\gamma_{b}}, \quad k_{ob} = \frac{\eta_{oo}}{2\gamma_{ob}}.\]
Then, the ratio $\overline{\gamma} = \gamma_o/\gamma_b$ defines the relative softness between the two cell types. Without loss of generality, we assume that $\overline{\gamma} \geq 1$ (i.e. the orange cells are relatively harder). Finally, the ratio $\overline{\eta} = \eta_{ob}/\eta_{oo}$ defines the relative repulsion strength between orange and blue cells and between orange cells. 

From now on, we take the blue cells as a reference and set $\gamma_{b}=1$ and $k_b = k_{ob} = 0.4$. We vary the three remaining ratios $\overline{\eta}$, $\overline{\gamma}$ and $\overline{k} = k_o/k_b$ which conveniently partitions the phase space into six regions (labeled A to F in Fig.~\ref{fig:3Dsorting}A,B) obtained by permuting the ordering of the repulsion strengths $\eta_{oo}, \eta_{ob}, \eta_{bb}$. They are depicted in the two phase diagrams in the $(\overline{k},\overline{\gamma}^{-1})$-plane for $\overline{\eta} > 1$ (i.e. the orange cells have a stronger affinity with their own type, Fig.~\ref{fig:3Dsorting}A) and $\overline{\eta}<1$ (i.e. the orange cells have a stronger affinity with the other type, Fig.~\ref{fig:3Dsorting}B). 

The results show that our model can reproduce at least all the configurations described in \cite{armstrong_cell_1989,brodland_differential_2002} namely: homophilic sorting with total or partial engulfment of one of the populations (resp. A, D and B, E) , homophilic sorting with separation by the medium (C,F), heterophilic/checkerboard sorting (D,E when $\overline{\gamma}$ tends to 1). More precisely, when $\overline{\eta}>1$, decreasing the compaction of the harder particles (i.e. increasing $\overline{k}$) makes the cell aggregate evolve from total engulfment (A) to partial engulfment (B) and separation (C), until the total dissociation of the aggregate by the medium when $k_1>1$ (Fig.~\ref{fig:3Dsorting}A). In particular, in the region A, the orange aggregate is highly compact but the repulsion force between orange and blue cells is relatively lower ($\eta_{ob}<\eta_{oo}$) which drives progressive engulfment, even from an initial segregated configuration (Fig.~\ref{fig:3Dsorting}C) as in \cite{brodland_differential_2002,mohammad_numerical_2022}. When $\overline{\eta}<1$ transitions are better visualized along the $1/\overline{\gamma}$ axis. For similar cells ($\overline{\gamma}=1$ and $\overline{k}<1/\overline{\eta}$), the heterophilic interactions ($\eta_{ob}<\min(\eta_{oo},\eta_{bb})$) produces a checkerboard patterning. Increasing the relative hardness of the orange cells is equivalent in our model to increase the repulsion strength $\eta_{oo}$ and decrease $\eta_{bb}$ which eventually leads from an engulfed/checkerboard state (D) to partial engulfment (E) and separation (F). Decreasing the compaction of the orange cells (i.e. increasing $\overline{k}$) prevents engulfment as it could be expected (Fig.~\ref{fig:3Dsorting}B). Representative situations are shown in the SM Videos~\ref{video:st_separation},\ref{video:st_checkerboard},\ref{video:st_internalization}, \ref{video:st_engulfment}. 

In the previous example, the cells have equal volume so their interface is always a plane and the interfacial terms with $\gamma_{ob}$ vanish (curvature $\kappa=0$). In the supplementary Fig.~{\ref{fig:compaction}} we show the equilibrium configuration for a pair of Laguerre cells with $\gamma_{ob}=1\ne0$ and different values of the volume ratio and compaction parameter ${k}:=\frac{\eta_{ob}}{2\gamma_o}$. Here we assume that larger particles are relatively softer (which comes from the definition of the cost function that prescribes the geometry of the cells and may be understood as a consequence of Laplace pressure although it is not explicitly modeled here). As a consequence, the interface with a smaller particle is curved. When ${k}$ is small, the cells have a stronger affinity. When the cells are identical (Fig~\ref{fig:compaction} right column), the cells may eventually merge into one spherical cell following Young-Dupr\'e relation. When the volume ratio decreases (left column), the smaller cell may eventually be internalized. These results are qualitatively consistent with the model in \cite{maitre_asymmetric_2016,runser_simucell3d_2024}. Note that we do not seek a quantitative comparison here as our system is not based on the same energy minimization formulation; moreover, an extended version of Young-Dupr\'e relation to this case would depend on the physical properties and modeling choices for nonidentical particles with different volumes. As an illustrative example with the simple modeling choices adopted here, we show a visually realistic simulation of bubbles in the Supplementary Material (Fig~\ref{fig:bubbles}).

\section*{Discussion}

We have introduced a novel approach to model, within a same framework, diverse systems ranging from individual deformable cells to complex tissue aggregates. The core idea is the notion of Laguerre tessellation in connection with optimal transport theory. Our model supports arbitrary shapes and physical deformations with a strict volume constraint, allowing complex soft-body simulations, with applications in particular in computational biology. Although our method is formally an extension of the idea of Voronoi tessellation, we have highlighted links with other independent methods such as point-particle systems, level-set methods and DCM. We also provide an efficient GPU implementation which leverages optimal transport techniques. 

Since, to the best of our knowledge, our approach is new in the literature, the main focus of the present article was to validate it by comparison to other classical computational models, thus providing an independent validation of their conclusions. The next important step is to seek a direct biological validation of our method by the confrontation to experimental data: a first step would be to fit Laguerre tessellations to 3D imaging data, as recently done in \cite{bourne_inverting_2024} for 2D material science data. Then, our method seems well suited to the in silico modeling of the development of blastocysts. Recent experimental findings \cite{ichikawa_ex_2022,belousov_when_2023} have identified several complex phenomena with morphological changes, sorting phenomena and dynamical topological properties via the formation of lumen. In order to understand the regulation processes between these phenomena, we need a mathematical model that can handle both individual cell deformations and their behaviors among dense aggregates. This seems particularly challenging using traditional methods, or at the price of an unreasonable computational cost, especially in a three-dimensional setting. This biological question was the initial motivation for the development of our method and is still an on-going work.

On the theoretical and numerical sides, several open questions and future improvements could be considered:
\begin{enumerate}
    \item Since our model has an intrinsic point-particle description, a natural question is to extend the active vertex theory developed in \cite{barton_active_2017,bi_motilitydriven_2016,sulsky_model_1984} to the case of Laguerre tessellations with volume constraints. In the same vein, it would be desirable to study whether the interaction mechanisms that we have introduced (or new ones) can be recast into a more common energy minimization framework.  The main theoretical and computational difficulties for these endeavors are linked to the differentiability properties of the solutions of semi-discrete optimal transport problems.
    \item A clear advantage of point-particle descriptions is the potential to derive coarse-grained continuum models better suited to mathematical analysis \cite{buttenschon_bridging_2020}. For particles with a volume, this is, to the best of ou knowledge, a largely open question, with the notable exception of \cite{santambrogio_optimal_2015,natale_gradient_2023} in the context of crowd evacuation modeling. A preliminary formal analysis (Supplementary Material) suggests a methodology based on mean-field theory that should lead to a novel continuum approach for soft-body systems.
    \item Optimal transport theory is historically linked to the design of numerical schemes for fluid flows. Thus, it provides a direct way to include fluid-structure interactions between macroscopic cells immersed in a fluid environment, in the mold of  \cite{peskin_immersed_2002,levy_partial_2022}. In particular, a discretization of the medium using \cite{gallouet_lagrangian_2018,levy_partial_2022} could address the momentum conservation issues that result from the simplification of the medium as a zero-cost Laguerre cell. 
    \item Although our GPU implementation is already quite efficient and operational for most cases, performance improvements could be considered. In particular, complex dynamical costs may remain challenging to handle, especially when they are defined implicitly as in level-set methods \cite{yang_modeling_2008} due to our symbolic computation framework \cite{charlier_kernel_2021}. Moreover surface interactions currently rely on isocontouring algorithms that would benefit from a fully GPU implementation for very large systems. Based on current benchmarking results, these improvements could lead to the development of a real-time 3D simulation tool for computational physics and biology. 
\end{enumerate}


\section*{Materials and Methods}

\subsection*{Implementation}

Our implementation framework is structurally similar to the so-called active vertex models \cite{barton_active_2017,bi_motilitydriven_2016}, simply looping back and forth between the computation of the space tessellation and the force terms. 

Starting from a set of locations, volumes and cost functions, we first compute an initial Laguerre tessellation. A classical procedure to ensure realistic initial configurations is known as Lloyd algorithm \cite{xin_centroidal_2016,bourne_laguerre_2020}. Then, the general equation~\eqref{eq:generalequation} is discretized in time using an explicit Euler(-Maruyama) scheme with a small time-step $\Delta t$. At each time step we use the current Laguerre tessellation to compute all the force terms needed to update the locations, volumes and costs. These new parameters are then used to compute the new current Laguerre tessellation. 

The most expensive part is the computation at each time step of a new Laguerre tessellation. We use a fast optimal transport solver implemented on the GPU which relies on a fine space discretization of the domain (Figs.~\ref{fig:graphical_abstract}d,\ref{fig:semidiscrete_d1}) as explained below.

The full code, with an API, a documentation and a gallery of examples can be found on GitHub and at \url{https://iceshot.readthedocs.io/}

\subsection*{Optimal Transport solver}

Whether it is in computational fluid dynamics or in materials science, the semi-discrete optimal transport problem is considered mostly for the squared Euclidean cost $c({x}, {y})$ 
$=|{x}-{y}|^2$. Extremely efficient numerical methods, based on a damped Newton algorithm, have been developed for this case and are now available as open source softwares (see for instance the packages \textsc{Geogram} or \textsc{Pysdot} available on GitHub). Unfortunately, these implementations leverage the structure of the squared Euclidean distance to compute Laguerre cells efficiently and cannot be applied with general cost functions.

In our approach, the freedom in the choice of arbitrary, individualized cost functions $c_i$ is of upmost importance. To unlock the use of Laguerre cells with general shapes and deformability properties, we work in a fully discrete setting with a discrete approximation of the domain $\Omega=[0,1]^d$ by a uniform grid of $M$ ``voxel centers'' ${y}_j$ (Fig.~\ref{fig:graphical_abstract}D. Non-square domains are simply cropped versions of $[0,1]^d$. We choose $M\gg N$ so that each particle is effectively discretized with a sufficiently high number of voxels ${y}_j$ with volume $1/M$. Solving discrete optimal transport problems is by now classical owing to numerous applications in data science \cite{peyre_computational_2019}. A quite reliable and stable method is the symmetrized and annealed variant of the renowned Sinkhorn algorithm \cite{feydy_geometric_2020}. However, as we target a tight fit to the volume constraints and since the dynamical setting naturally lets us re-use optimal potential values $w_i$ from one time step to another, we rely instead on a solver that leverages the semi-discrete structure of the problem. The main observation, proved for instance in \cite{bourne_semidiscrete_2018,xin_centroidal_2016}, is that the dual potentials $w=(w_i)_{i=1,\ldots,N}$ that appear in the definition \eqref{eq:Laguerre} of the Laguerre cells solve the dual maximization problem 
\[w = \argmax_{(w_i)_{i=1,\ldots,N}} \mathcal{K}(w),\quad \mathcal{K}(w) :=  \sum_{i=1}^N v_i w_i + \int_{{y}\in\Omega}\, \min_{i=1,\ldots,N} [c({y},{x}_i) - w_i]\, \dd {y}~.\]
The objective function $\mathcal{K}(w)$ is concave as the sum and minimum of functions that are linear with respect to $w$. Its gradient simply reads
\[\nabla \mathcal{K}(w) = \big(\,v_i - |\mathscr{L}_i|\,\big)_{i=1,\ldots,N}\]
where $\mathscr{L}_i$ is defined by \eqref{eq:Laguerre} for an arbitrary set of potentials.
This implies that the optimality conditions $|\mathscr{L}_i|=v_i$ are satisfied at the maximum of $\mathcal{K}$, which is well-defined if the sum of the target volumes $v_i$ does not exceed that of the domain~$\Omega$ and if the cost functions $c_i : {y} \mapsto c({y},{x}_i)$ satisfy mild technical assumptions: notably, distinct indices $i\neq j$ should correspond to distinct cost functions $c_i \neq c_j$.

In order to solve this maximization problem, as in \cite{buze_anisotropic_2024}, we rely on a classical quasi-Newton method, namely the L-BFGS-B algorithm. At each time step, the stopping criterion for this optimization algorithm is when the average volume deviation (between the computed volumes and the theoretical volumes) falls below 1\%. Except for rare isolated events, this is always the case in all the experiments presented. 

In a discretized domain $\Omega=\{{y}_1,\ldots,{y}_M\}$ with $M$ large, the most expensive part of the computation of the objective function is the minimum operation over the $i$-indices of the $N\times M$ matrix $(c({y}_j,{x}_i) - w_i)_{i,j}$ followed by a sum over the $j$-indices approximating the integral. In order to handle arbitrary cost functions and fine discretizations with $M\sim 10^6$ voxels, we rely on the semi-symbolic lazy tensor framework introduced by the \textsc{KeOps} and \textsc{GeomLoss} libraries \cite{charlier_kernel_2021}. This lets our plain Python code perform scalable geometric computations on the GPU, with automatic differentiation and without memory overflows.

The assignment of each voxel to one of the ${x}_i$ is then given by the following map, which in optimal transport corresponds to a discretized variant of the Monge transport map \cite{peyre_computational_2019}:
\[T : \{{y}_j\}_{j\in\{1,\ldots,M\}} \to \{{x}_i\}_{i\in\{1,\ldots,N\}},\quad {y}_j \mapsto {x}_{j^*},\]
with 
\[j^* := \argmin_{i\in\{1,\ldots,N\}}\{c({y}_j,{x}_i)-w_i\}.\]
Semi-discrete optimal transport on a discretized grid and the voxel assignment procedure is illustrated is dimension 1 in Fig.~\ref{fig:semidiscrete_d1}.

\subsection*{Software and Hardware}
 All the simulations presented in this article run on a Nvidia RTX A6000 GPU card. We use \textsc{NumPy}, \textsc{PyTorch}, \textsc{KeOps} for our simulations while relying on \textsc{Matplotlib}, \textsc{PyVista} and \textsc{Paraview} for visualizations.

\subsection*{Cost normalization}

We assume that the domain denoted by $\Omega$ is connected and bounded, with total volume normalized to 1 and typical length $L=1$. The cell positions ${x}_i$ and the other quantities are assumed to be properly adimensionalized with respect to this length scale.

Unless otherwise specified, the cost function $c$ is implicitly normalized so that, given a shape $\mathscr{S}_0\subset \Omega$ corresponding to a level set of $c$, with centroid at ${x}_0$, and a normalizing factor denoted by $\lambda$ in the parameter table,
\[\int_{\mathscr{S}_0} c({x},{x}_0)\dd{x} = \lambda |\mathscr{S}_0|.\]
In other words, the total cost of the shape $\mathscr{S}_0$ is a fraction of its volume. When $c$ is the power cost $c({x},{y}) = \lambda_\alpha^{-1}|{x}-{y}|^\alpha$, we consider a fixed radius $R_0$ and thus take in dimension $d$, 
\[\lambda_\alpha = \frac{\lambda^{-1} d}{\alpha+d}R_0^\alpha.\]
In this case, we always take $R_0$ as the average radius of the particles. For the anisotropic costs (ellipsoidal and spherocylinder costs), the cost of a particle with aspect ratio $r$ is normalized to be equal to a fraction $\lambda$ of the volume of the same shape with aspect ratio $r$ associated to a given small axis denoted by $b$ in the parameter table.

With this normalization choice, the incompressibility force defined \eqref{eq:incompressibilityforce} scales like $|\nabla_{{x}_i} \mathcal{T}_c(\hat{\mu})|\propto R_0$. We thus implicitly normalize the gradient descent step in Eq.~\eqref{eq:incompressibilityforce} as $\tau \equiv \tau/R_0$ so that the gradient descent motion is independent of the size $R_0$ of the particles. In other words, if one reduces or increases the volume of the particles, one simulates the same system but in a respectively larger or smaller domain. 

\subsection*{Boundary conditions}

In all experiments, there is either no boundary conditions or periodic boundary conditions. Without boundary conditions, we note that the only important quantity to consider is the cost function $c(x,y)=c(y-x)$ which is always a function of the connecting vector $y-x$ between two points $x,y\in\Omega=[0,1]^d$. In the periodic case, we implicitly replace this vector by its ``periodized'' version obtained by taking the shortest connecting vector between two representative points of $x$ and $y$ in the full space. More precisely, by denoting $z=(z^1,\ldots,z^d)$ the components of a vector $z$, its periodized version can be obtained by applying the following transformation on each component $z^k$: 
\[z^k \mapsto z^k + \chi_{ z^k<-1/2} - \chi_{z^k>1/2},\]
where $\chi_{z^k<-1/2}=1$ if $z^k<-1/2$ and $0$ otherwise. To alleviate notations, we only write the cost function and other quantities assuming no boundary conditions but we implicitly use the periodized version of the vectors to handle periodic boundary conditions. Besides, positions are simply computed modulo 1 when needed.

\subsection*{Anisotropic shapes}

The Principal Component Analysis of an anisotropic shape $\mathscr{L}_i$ is computed in a discretized setting as the empirical covariance matrix 
\[\hat{\Sigma}_i = \frac{1}{N_i - 1}\sum_{{y}\in \mathscr{L}_i} ({y} - {x}_i)({y} - {x}_i)^\mathrm{T},\quad N_i := \#\{{y}\in \mathscr{L}_i\}.\]
The covariance matrix defining the orientation of a particle in Figs.~\ref{fig:benchmark}c,\ref{fig:shapes}b is simply defined as 
\[\Sigma_i = \frac{1}{\det(\hat{\Sigma}_i)^{1/d}}\hat{\Sigma}_i.\]
The direction of motion ${n}_i$ is defined as the normalized eigenvector associated to the largest eigenvalue of the matrix $\hat{\Sigma}_i$ (with the sign correctly chosen to avoid flipping the direction of motion). 

\subsection*{Surface meshing and curvature computation}

The computation of interfacial forces \eqref{eq:interfaceforce} requires to integrate quantities along the boundary of each Laguerre cell. Since our optimal transport algorithm outputs volumetric data, we use the fast isocontouring algorithm SurfaceNets \cite{schroeder_highperformance_2024} which is integrated in VTK \cite{schroeder_visualization_2006} to convert at each time step the Laguerre cells into mesh data. We then rely on the routines already implemented in VTK and PyVista \cite{sullivan_pyvista_2019} to compute the normal and curvature of each mesh element. For better numerical stability we filter the mesh elements for which the normal direction is not well defined.


\clearpage 

%
\printbibliography

%
%
%
%
%
%


\section*{Acknowledgments}
This article was inspired by discussions with Takafumi Ichikawa on the modeling of embryonic development and with David Bourne, Maciej Buze and Steven Roper on the modeling of polycrystalline metals.
The authors wish to thank Bruno Lévy, Roya Mohayaee, Steffen Plunder, Tsubasa Sukekawa, Tetsuya Hiraiwa for their insightful advice and for the useful discussions. 
\paragraph*{Funding:}
The work of AD is supported by the following grant: KAKENHI Grant-in-Aid for Early-Career Scientists (Grant number 23K13015). The work of JF is supported by the French ``Agence Nationale de la Recherche'' via the ``PR[AI]RIE-PSAI'' project (ANR-23-IACL-0008). 
\paragraph*{Author contributions:}
Conceptualization: AD and JF.
Methodology: AD and JF.
Software: AD and JF.
Validation: AD and JF.
Formal analysis: AD and JF.
Investigation: AD and JF.
Resources: AD and JF.
Data curation: AD and JF.
Writing - original draft: AD and JF.
Writing - review \& editing: AD and JF.
Visualization: AD and JF.
Supervision: AD and JF.
Project administration: AD and JF.
Funding acquisition: AD and JF.
\paragraph*{Competing interests:}
All authors declare no competing interests.
\paragraph*{Data and materials availability:} This study did not generate new materials. The full code (including the parameter sets) to reproduce all the simulations is available on GitHub at \url{www.github.com/antoinediez/ICeShOT} and on the Zenodo archive \url{https://doi.org/10.5281/zenodo.18603820}. The associated documentation website \url{https://iceshot.readthedocs.io/} includes all the supplementary videos mentioned in the present article.

\newpage


\begin{figure}
    \centering
    \includegraphics[width=0.95\textwidth]{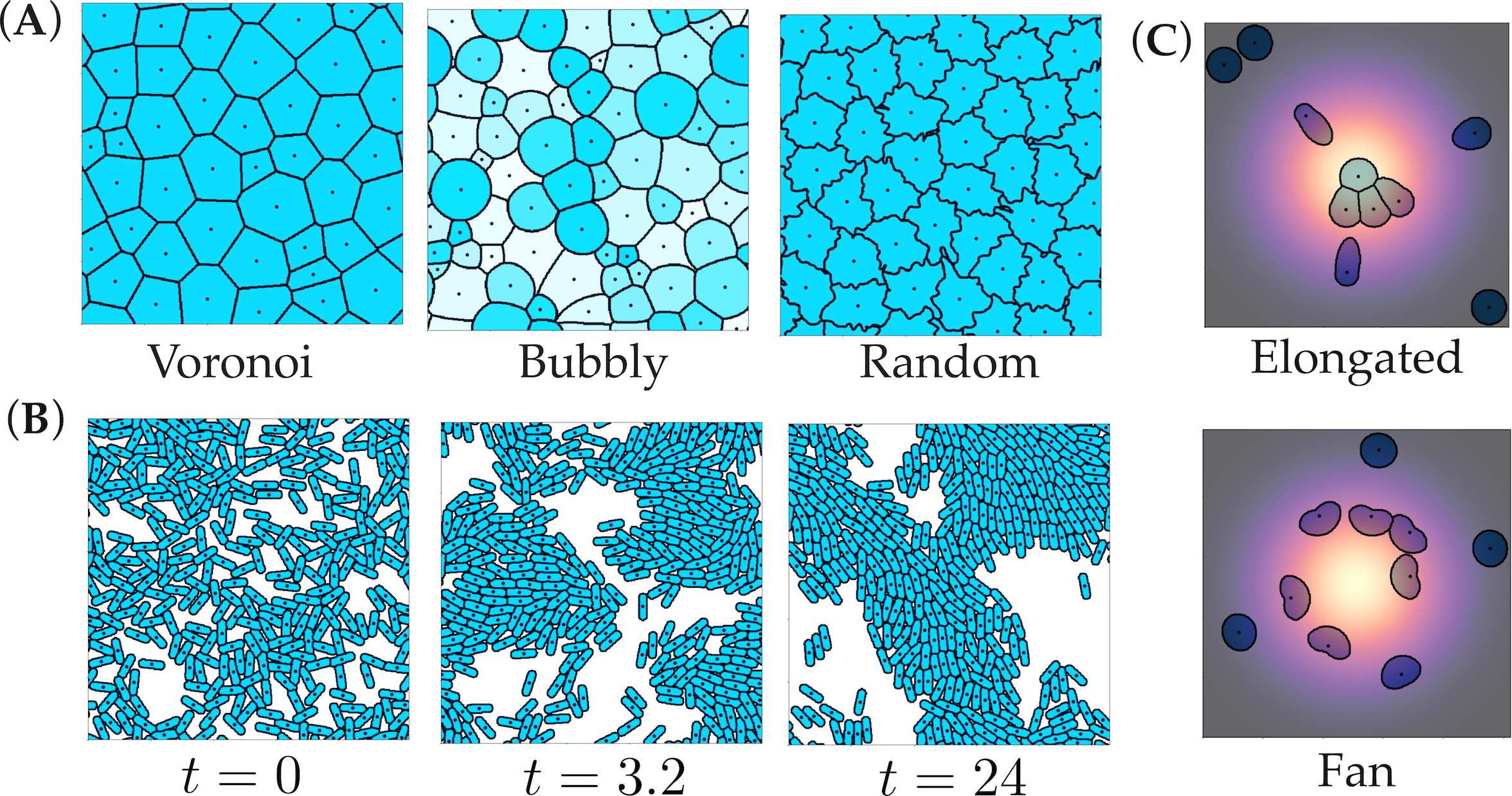}
    \caption{\textbf{Static and dynamic shapes in 2D.} \textbf{(A)} Three Laguerre tessellations: (Left) Voronoi diagram  obtained with the $L^2$ cost and random volumes $v_i$ sampled uniformly with a ratio 1/5. (Center) A bubbly tessellation similar to \cite{ishimoto_bubbly_2014} obtained with 66 particles with random volumes sampled uniformly with a ratio 1/20 and power costs \eqref{eq:powercost} with exponents distributed uniformly between 0.5 and 4. Lighter colors indicate lower values of this exponent and correspond to softer shapes. (Right) Voronoi tessellation with random fluctuations similar to \cite{miyazaki_mechanism_2023} obtained with 42 identical particles and a randomly perturbed $L^2$ cost. \textbf{(B)} A swarm of rod-shaped particles obtained with the cost \eqref{eq:spherocylindercost} showing the emergence of long-range alignment due to deformations and non-overlapping interactions. See also SM Video~\ref{video:rodshape}. \textbf{(C)} Chemotaxis motion induced by shape deformations with two choices of biased costs \eqref{eq:costpotentialchemotaxis}-\eqref{eq:bias} See also SM Videos~\ref{video:chemo_long},\ref{video:chemo_fan}}
    \label{fig:shapes}
\end{figure}

\begin{figure}
    \centering
    \includegraphics[width=0.95\textwidth]{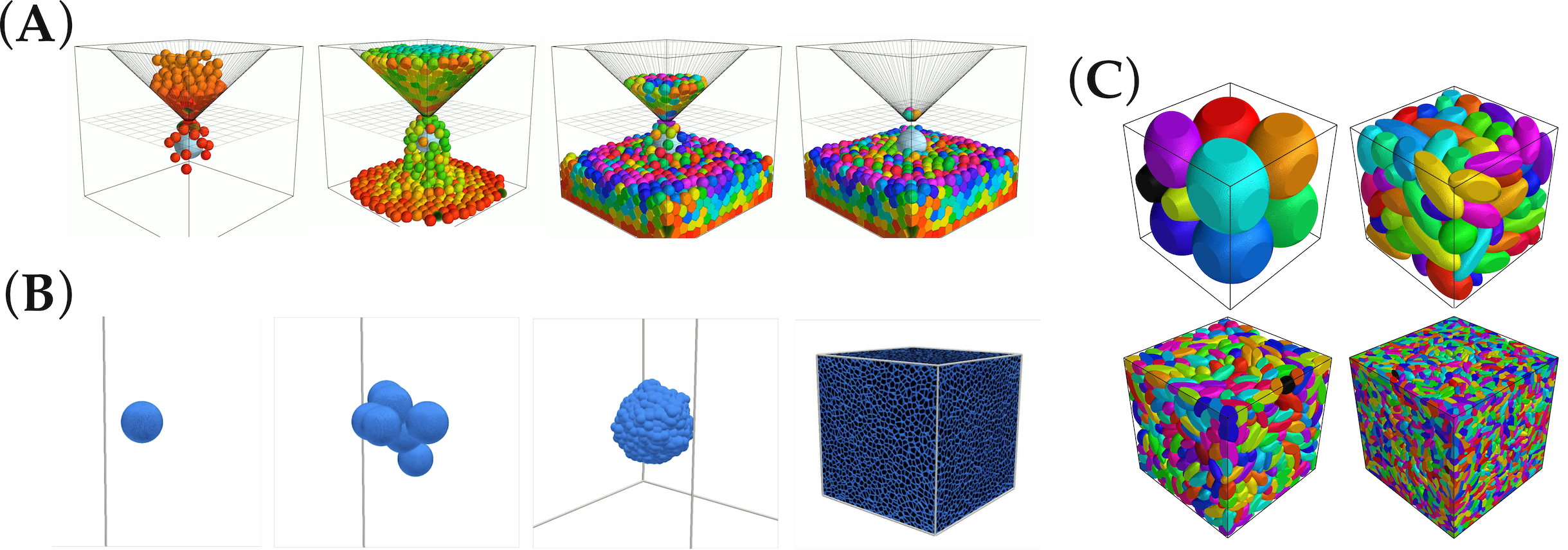}
    \caption{\textbf{3D benchmarking examples.} \textbf{(A)} Falling soft-spheres in a hourglass domain. See also Supplementary Video~\ref{video:fallingstuff}. \textbf{(B)} Exponential growth of a 3D aggregate via successive cell division and growth phases, zooming out from $N=1$ to $N=50,000$ cells. See also SM Video~\ref{video:tissue_growth_3D}. \textbf{(C)} Final configuration of a deformation-driven run-and-tumble motion for $N=10,100,1000,10000$ deformable ellipsoids with a space discretization grid of size $M=512$. See also SM Video~\ref{video:benchmark}.}
    \label{fig:benchmark}
\end{figure}

\begin{figure}
    \centering
    \includegraphics[width=0.99\textwidth]{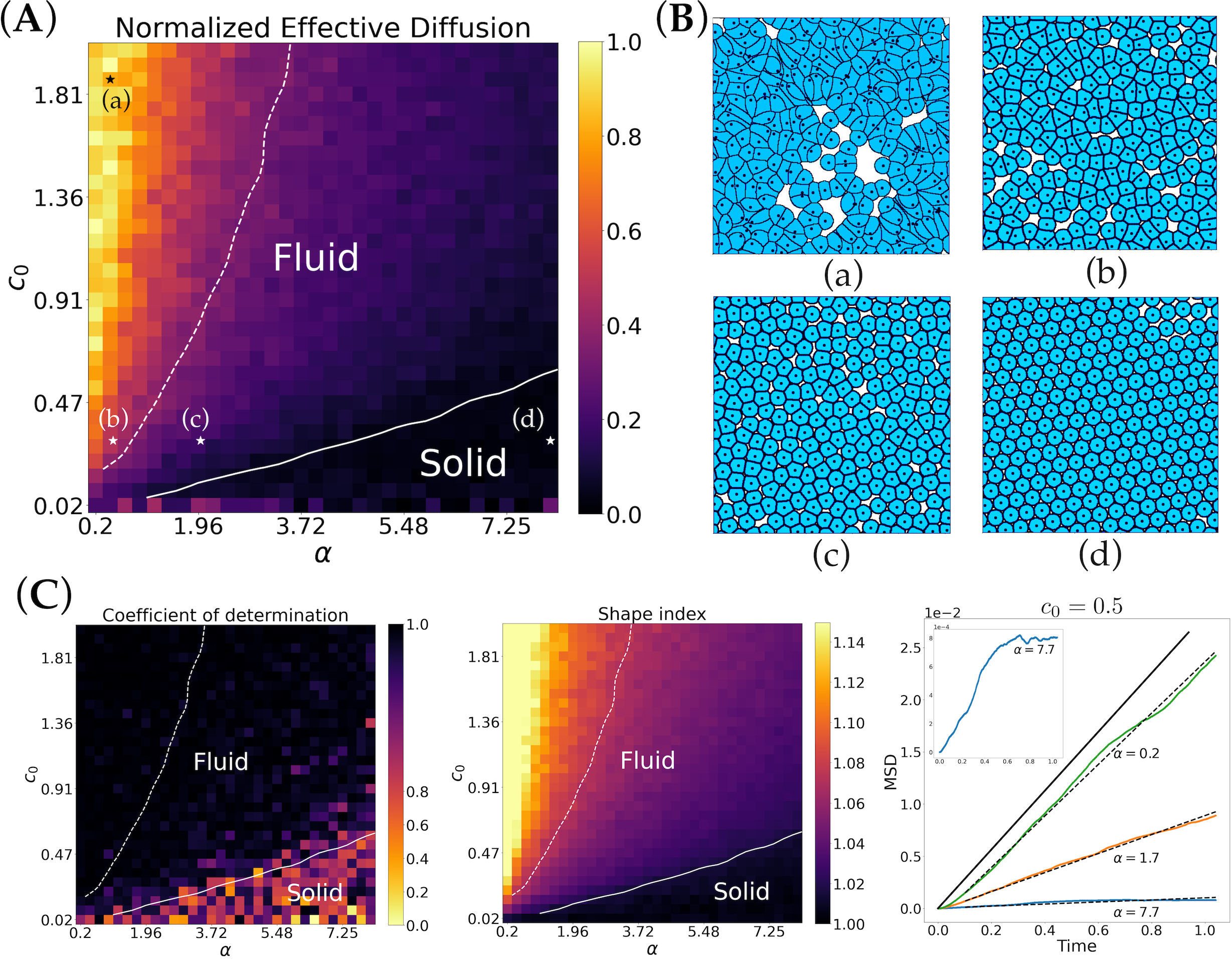}
    \caption{\textbf{Active Brownian Particles with deformations.} \textbf{(A)} Phase diagram in the $(\alpha,c_0)$ plane. \textbf{(B)} Snapshots of the final configuration for different values of $(\alpha,c_0)$ indicated by the labels (a),(b),(c),(d) on the phase diagram. See also SM Videos~\ref{video:abp_a05},\ref{video:abp_a2},\ref{video:abp_a10}. \textbf{(C)} (left) Coefficient of determination of the linear fit in the $(\alpha,c_0)$ plane. The center of the color map is set at $r^2=0.9$. (center) Average shape index $\langle\sigma_i\rangle$ in the $(\alpha,c_0)$ plane. The center of the colormap is set at $\sigma=3.81/(2\sqrt{\pi})$. (right) MSD (plain line) and linear fit (dashed line) over time for $c_0=0.5$ and $\alpha=0.2, 1.7, 7.7$. The black line has slope $\sigma_\text{thr}$. Insert: zoom on the MSD curve $\alpha=7.7$ showing saturation when the system has reached a stable hexagonal pattern. See also Fig.~\ref{fig:msdloglog}A for the same plot but over a longer time interval.}
    \label{fig:abp}
\end{figure}

\begin{figure}
    \centering
    \includegraphics[width=0.98\textwidth]{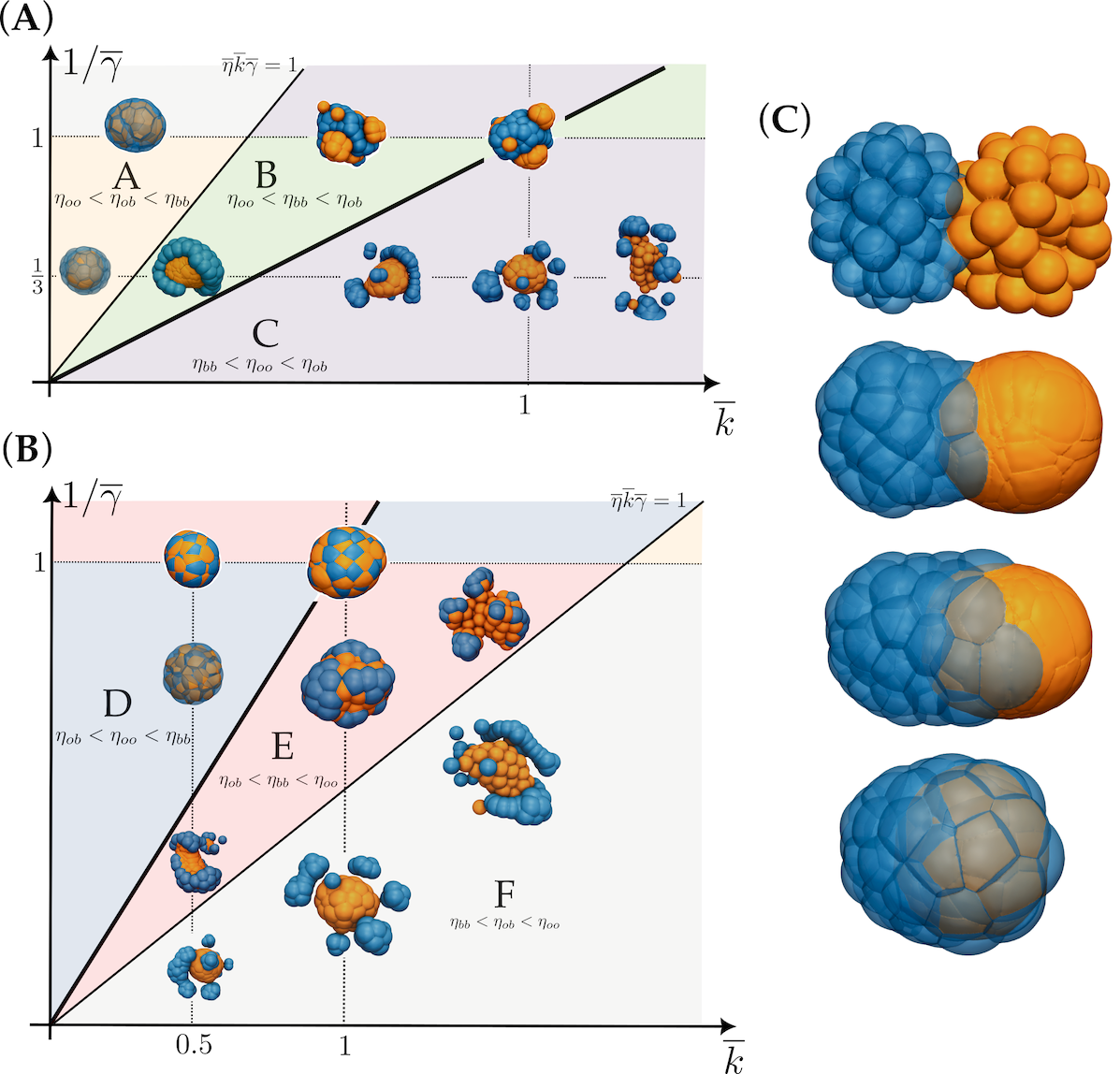}
    \caption{\textbf{Sorting patterns in 3D.} \textbf{(A)}-\textbf{(B)} Sorting patterns in a homogeneous mixture of two cell types with resp. $\eta=3>1$ and $\eta=0.3<1$, obtained from an initial mixed aggregate of $N=120$ cells, under different conditions on the surface tensions parameters. The phase domain compares relative compactness $\overline{k}$ with relative softness $\overline{\gamma}$. The line $\{\overline{k}\overline{\gamma}=1\}$ defines the boundary between the regions $\{\eta_{oo} \lessgtr \eta_{bb}\}$ and the line $\{\overline{k}\overline{\gamma}\overline{\eta}=1\}$ defines the boundary between the regions $\{\eta_{ob}\lessgtr\eta_{bb}\}$. Two disconnected regions with the same background color are equivalent with blue and orange cells inverted depending on $\overline{\gamma}\gtrless1$. \textbf{(C)} Engulfment of an aggregate of orange cells in an aggregate of blue cells starting from a configuration where the two cell types are separated (region A: $\overline{\eta}=3$, $\overline{\gamma}=2$, $\overline{k}\overline{\gamma}\overline{\eta}=0.8$), see also SM Video~\ref{video:st_engulfment}.}
    \label{fig:3Dsorting}
\end{figure}

\clearpage
\begin{table}
    \centering
    \caption{\textbf{Benchmark.} Total computation time for 2000 iterations and various values of $N$ and $M$ in dimension $d=3$.} 
    \begin{tabular}{c||cccc}
        $N$ $\backslash$ $M$ & 64      & 128      & 256   & 512 \\\hline\hline
        10                      & 30 s & 1.3 min  & 7.4 min & 62 min\\
        100                     & 45 s & 1 min  & 4.7 min & 36 min\\
        1000                    & NR      & 4 min & 25 min & 3 h\\
        10000                   & NR      & NR       & 4.9 h & 35 h \\
    \end{tabular}
        \label{tab:benchmark}
\end{table}


\subsection*{Supplementary materials}
Supplementary Text\\
Supplementary Figures \ref{fig:semidiscrete_d1} to \ref{fig:compaction}\\
Tables \ref{tab:parameters} and \ref{tab:parameters_supplementary}\\
Movies \ref{video:incompressibility} to \ref{video:bubbles}\\


\newpage


\renewcommand{\thefigure}{S\arabic{figure}}
\renewcommand{\thetable}{S\arabic{table}}
\renewcommand{\theequation}{S\arabic{equation}}
\renewcommand{\thepage}{S\arabic{page}}
\setcounter{figure}{0}
\setcounter{table}{0}
\setcounter{equation}{0}
\setcounter{page}{1} 


\begin{center}
\section*{Supplementary Materials for\\ \scititle}

Antoine~Diez$^{\ast\dagger}$,
J.~Feydy$^\dagger$,\\
\small$^\ast$Corresponding author. Email: antoine.diez@riken.jp\\
\small$^\dagger$These authors contributed equally to this work.
\end{center}

\subsubsection*{This PDF file includes:}
\begin{figure}
    \centering
    \includegraphics[width=0.8\linewidth]{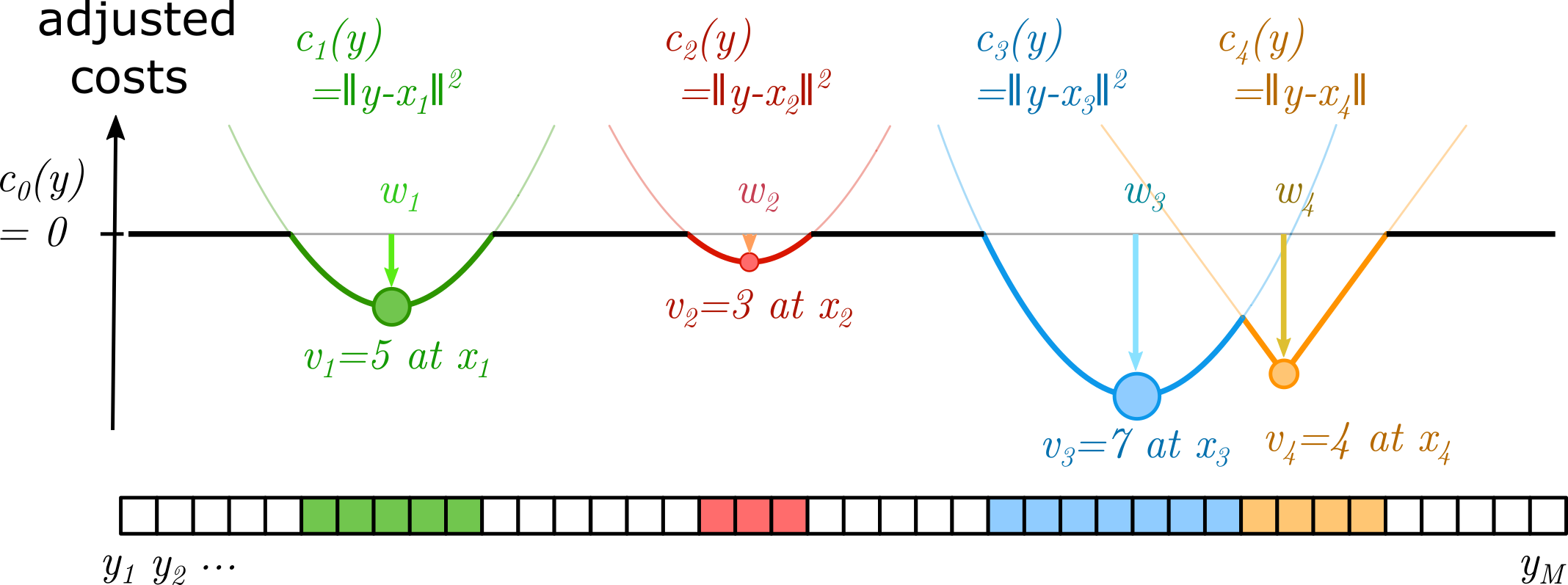}
    \caption{\textbf{Semi-discrete optimal transport on a grid in dimension 1 with four particles.} Each voxel is assigned to the minimal adjusted cost. The potentials $w_i$ are adjusted to satisfy the volume constraints.}
    \label{fig:semidiscrete_d1}
\end{figure}

Supplementary Text\\
Supplementary Figures \ref{fig:semidiscrete_d1} to \ref{fig:compaction}\\
Tables \ref{tab:parameters} and \ref{tab:parameters_supplementary}\\
Captions for Movies \ref{video:incompressibility} to \ref{video:bubbles}\\

\subsubsection*{Other Supplementary Materials for this manuscript:}
Movies \ref{video:incompressibility} to \ref{video:bubbles}\\

\newpage


\subsection*{Supplementary text}

\subsubsection*{Supplementary experiments}

The transition from a fluid to a solid phase as shown in the main text plays an important role in the study of congestion effects in crowd modeling. Here, the active Brownian particle model can be adapted to an evacuation situation where all particles (modeling pedestrians) move towards a single exit point. When the pedestrians are assumed to have the ability to deform ($\alpha=0.5$), the crowd behaves as a fluid and safely evacuates the room. When pedestrians are modeled by hard-spheres ($\alpha=8$), congestion effects appears, typically materialized by the formation of an arch around the exit point \cite{maury_handling_2011}, which is stable even in the presence of noise (Fig.~\ref{fig:evacuation} and SM Videos~\ref{video:crowd_a05},\ref{video:crowd_a8}).

In the main text, we consider a homogeneous population of active deformable particles with different deformability properties. It is also possible to mix soft and hard particles by assigning different costs to different populations. As an example, we consider a 2D falling soft-sphere experiment.  In a crowded situation, there is a competition between the incompressibility force $-\tau\nabla_{{x}_i}\mathcal{T}_c$ and the gravity-like force ${F}^\mathrm{point}_i = -F_0{e}_z$. When the population is homogeneous, we recover the results shown in the main text in particular a hard-sphere packing configuration. In this particular situation, when $\alpha$ is small, the particles manage to squeeze their way by adopting elongated columnar shapes (Fig.~\ref{fig:fallingspheres}).

To study the effect of heterogeneity we now consider the same situation but with two equal populations associated to two values $\alpha=1$ and $\alpha=2$. In addition, we also consider two different values for the gradient-step parameter $\tau$, which plays the role as an inertia parameter: although the particles with $\alpha=1$ are always softer, they appear as heavier when $\tau$ is small and lighter when $\tau$ is big. It leads to a sorting phenomena where lighter particles are pushed on top (Fig.~\ref{fig:fallingspheres_mixed} and SM Videos~\ref{video:fs_tau8},\ref{video:fs_tau3},\ref{video:fs_tau1}). It is always easier for softer particles to squeeze their way down, which leads to non-convex or elongated shapes.

Biological cells are often regarded as similar to soap bubbles. Laguerre tessellation have actually already been used to model bubbles and foams in computer graphics \cite{busaryev_animating_2012} but with a simpler model compared to the model presented in the main text. Compared to biological cells, bubbles are simpler because they must satisfy the Plateau conditions which impose a constant curvature and a $120^\circ$ angle between two bubbles of equal size. In our model, they can be enforced by setting all surface tension parameters to 1. Simulating bubble interactions is however challenging as they can have very heterogeneous volumes and their number is not conserved in time due to frequent fusion, explosion, creation events. In particular, traditional DCM would need to deal with many rearrangements and topological transitions. Our model does not has such flaws. As an illustrative example we simulate a very simple system of bubbles. We assume that bubbles are generated randomly within a fluid which occupies the bottom half of a square domain. Due to gravity and buoyancy forces, the bubbles stay at the interface between the fluid and the air. We add a small force which pushes the bubbles towards the boundary of the domain as in \cite{busaryev_animating_2012}. We assume that bubbles can explode randomly with a higher probability when they are larger. Finally, two bubbles can randomly fuse when they are in contact, which creates a single bubble with a volume which is the sum of the two fused bubbles. The simulation is shown in SM Video~\ref{video:bubbles} and Fig.~\ref{fig:bubbles}.

Table~\ref{tab:parameters_supplementary} gathers the modeling and numerical parameters for the supplementary experiments.

\subsubsection*{Optimal transport point of view}

We gather here some important definitions and results that may be useful for a better understanding of the main text. Optimal transport theory is by now well-established and we refer to \cite{santambrogio_optimal_2015} and the references therein for a deeper introduction to the subject from a mathematical perspective and to \cite{peyre_computational_2019,feydy_geometric_2020} for a focus on numerical methods and applications in data sciences and computational geometry. 

Historically, Monge considered the problem of transporting sand piles located at some positions ${y}_1,\ldots,{y}_M\in\Omega\subset \mathbb{R}^d$ and with respective masses $\beta_1,\ldots,\beta_M$, to some target positions ${x}_1,\ldots,{x}_N\in\Omega$ that should respectively receive a volume $\alpha_1,\ldots,\alpha_N$ of sand. The total volume $\sum_j \beta_j = \sum_i \alpha_i = 1$ is classically normalized to~1. The transport of a sand grain between the locations ${y}$ and ${x}$ comes with a cost $c({y},{x})$; the optimal transport problem is to find an optimal way of moving all of the sand at a minimal cost. When $N=M$ and $\alpha_i=\beta_j=1/N$, this problem can be seen as a \textit{matching} problem, which consists in finding the one-to-one map (or equivalently a permutation of the set $\{1,\ldots,N\}$)
\[T:\{{y}_1,\ldots,{y}_N\} \to \{{x}_1,\ldots,{x}_N\},\]
that minimizes the total transport cost
\[\mathcal{T}_c := \frac{1}{N}\sum_{i=1}^N c({y}_i,T({y}_i)).\]
In the general case with $N\ne M$ and arbitrary volumes, such a \textit{Monge map} may not exist since we may have to split the sand piles and send their different parts to different locations. The optimal transport problem then only makes sense for the more general notion of \textit{transport plan} introduced by Kantorovich. In this setting, a transport plan is a $N\times M$ matrix $\pi=(\pi_{ij})$ whose coefficients $\pi_{ij}~\geq0~$ indicate how much of the sand pile $j$ is sent to the location $i$. Consequently, the transport plan $\pi$ should satisfy the constraints 
\[\forall j,\,\,\sum_{i=1}^N \pi_{ij} = \beta_j,\quad \forall i,\,\,\sum_{j=1}^M \pi_{ij} = \alpha_i.\]
The optimal transport problem then consists in finding the matrix $\pi$ that satisfies this constraints and minimizes the total transport cost 
\[\mathcal{T}_c := \sum_{i=1}^N\sum_{j=1}^M  \pi_{ij} c({y}_j,{x}_i).\]
Going further, this problem can be reformulated using measure theory. The sand piles and target locations are respectively modeled by the discrete probability measures 
\begin{equation}\label{eq:discretealphabeta}\nu = \sum_{j=1}^M \beta_j \delta_{{y}_j},\quad \mu = \sum_{i=1}^N \alpha_i \delta_{{x}_i}.\end{equation}
The Monge problem corresponds to the minimization problem 
\begin{equation}\label{eq:Monge_sm}\inf_{T:\Omega\to \Omega} \left\{ \int_{\Omega} c({y},T({y})) \nu(\dd {y})\,\Big|\,\,\,T\#\nu=\mu\right\},\end{equation}
where we recall that the $\textit{push-forward}$ measure $T\#\nu$ is defined as the measure on $\Omega$ such that 
\[T\#\nu(\mathscr{B}) = \nu(T^{-1}(\mathscr{B})),\]
for all open sets $\mathscr{B}$. 

The Kantorovich problem corresponds to the minimization problem 
\begin{equation}\label{eq:Kantorovich_sm}\inf_{\Pi \in\mathcal{P}(\Omega\times\Omega)} \left\{ \int_{\Omega}\int_{\Omega} c({y},{x}) \Pi(\dd {y},\dd{x})\,\Big|\,\,\Pi_1 = \nu,\,\,\Pi_2 = \mu\right\},\end{equation}
where $\mathcal{P}(\Omega\times\Omega)$ denotes the set of probability measures on $\Omega\times\Omega$ and $\Pi_1$ and $\Pi_2$ are the first and second marginals of $\Pi\in \mathcal{P}(\Omega\times\Omega)$. 

Although it can be checked that with the choice \eqref{eq:discretealphabeta}, the Monge and Kantorovich problems reduce to the optimization problems introduced above, the formulations \eqref{eq:Monge_sm}-\eqref{eq:Kantorovich_sm} are much more general and are not restricted to the case of discrete measures $\mu,\nu$. In particular, $\mu$ and/or $\nu$ can be continuous measures (defined by their probability density function). In this case Brenier, Gangbo, McCann, Caffarelli and others have proved the following fundamental theorem. 
\begin{theorem}
    If the source measure $\nu$ is a continuous measure and under some assumptions on the cost function $c$, then the Monge problem \eqref{eq:Monge_sm} has a unique solution $T$. 
\end{theorem}

In our approach, we consider the semi-discrete case where $\mu=\sum_i \alpha_i \delta_{{x}_i}$ is a discrete measure and $\nu = \mathrm{Leb}$ is a continuous measure, here the Lebesgue measure. In this case, a direct application of the previous theorem shows that the optimal transport problem is equivalent to the computation of a partition of the space $\Omega=\cup_i \mathscr{S}_i$ into the disjoint sets: 
\[\mathscr{S}_i := T^{-1}(\{{x}_i\}).\]
Moreover, a simple computation shows that $\nu(\mathscr{S}_i) = \alpha_i$. The fundamental theorem of semi-discrete optimal transport theorem, which is the theoretical basis of our work, can be found in \cite{bourne_semidiscrete_2018,kitagawa_convergence_2019}. It proves that, for a large class of cost function, our problem always has a unique solution. 

\begin{theorem}\label{thm:semidiscrete}
Let $\hat{\mu} = \sum_{i=1}^N v_i \delta_{{x}_i}$ be a discrete probability measure on $\Omega$. Let $c$ be a cost function such that for all $i$, $c_i : {x}\in \Omega \mapsto c({x},{x}_i)\in [0,+\infty) $ belongs to $C^{1,1}(\Omega)$ and ${y}\mapsto \nabla_{{x}} c({x},{y})$ is injective for all ${x}\in\Omega$. Then the Monge problem \eqref{eq:Monge_sm} has a unique solution $T$ given by 
\[T : \Omega \to \{{x}_i\}_i,\,\, {x}\in \mathscr{L}_i \mapsto {x}_i\]
where the partioning sets $(\mathscr{L}_i)_i$ are the Laguerre cells defined by \eqref{eq:Laguerre}.
The intersection of two Laguerre cells has zero Lebesgue measure and the weights $w_i\in \mathbb{R}$, called Kantorovich potentials, are uniquely defined such that for all $i\in\{1,\ldots,N\}$, $|\mathscr{L}_i| = v_i$.

\end{theorem}

\begin{figure}
    \centering
    \includegraphics[width=0.9\textwidth]{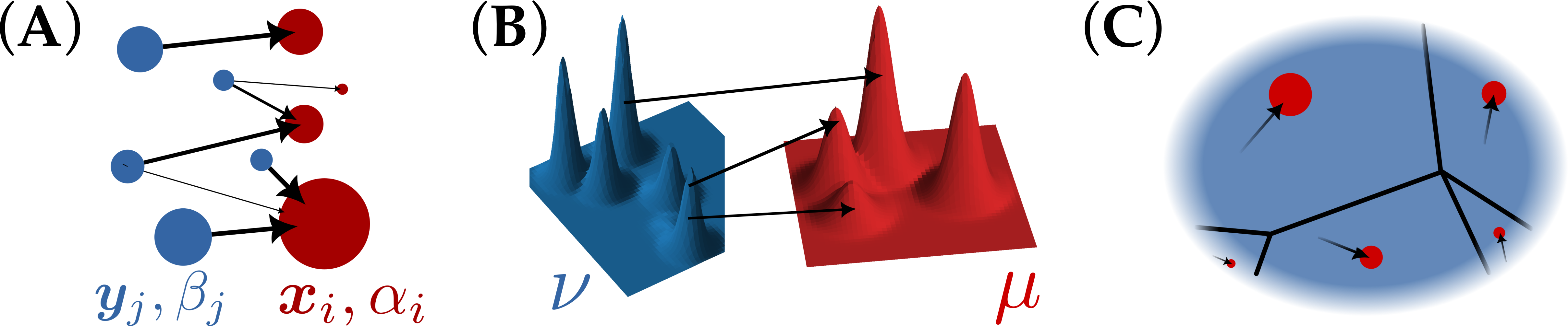}
    \caption{\textbf{Discrete, continuous and semi-discrete optimal transport.} \textbf{(A)} A discrete optimal transport problem is a matching point problem but which may require mass splitting in which case there is no Monge map. \textbf{(B)} If the source measure has a density, then there is a Monge map. \textbf{(C)} In the semi-discrete case, the target measure is discrete and the Monge map defines a partition of the space.}
    \label{fig:ot}
\end{figure}

A graphical illustration of the discrete, continuous and semi-discrete problems is shown in Fig.~\ref{fig:ot}.

\subsubsection*{Coarse-grained analysis}

There is a fundamental duality between agent-based and continuum Partial Differential Equations (PDE) models in mathematical biology. Each description has its own strengths and weaknesses. Typically, agent-based models provide the finest level of details and modeling freedom and are naturally adapted to in silico computations with potentially stochastic components. However, they provide little room for theoretical mathematical analysis due to their complexity and usually high-dimensional nature. On the contrary, continuum PDE models provide a synoptic point of view at a statistical scale, benefit from all the analytical mathematical machinery, are often closer to physical laws and can eventually lead to much stronger ``proved'' conclusions. Bridging the gap between the two descriptions in a rigorous mathematical manner is a fundamental issue in mathematical physics and mathematical biology, but its feasibility strongly depends on the modeling framework \cite{buttenschon_bridging_2020}. Point-particle systems are classically the easier to coarse-grain, as they can often be written in a statistical physics framework for which many theoretical results are now available \cite{chaintron_propagation_2022}. Particles with a shape, regardless of the description (vertex models, cellular automata, phase-fields etc), seem much more difficult to handle, since no universal or natural continuum limit and scaling can be easily deduced. The framework introduced in the present article is somehow hybrid, it can be seen as an intrinsically agent-based model~\eqref{eq:generalequation} but allows a control on the shape and volume exclusion parameters. Formal computations (at this stage) and earlier works in computational optimal transport suggest that this underlying point-particle description can naturally lead to coarse-grained continuum PDE models.

Let us consider a simplified version of the equations of motion \eqref{eq:generalequation}, with no surface interaction and independent Brownian noises:
\begin{equation}\label{eq:meanfieldincompressible_sm}
    \dd {x}_i = {b}({x}_i,\hat{\mu})\dd t - \tau \nabla_{{x}_i} \mathcal{T}_c\dd t  + \sqrt{2\sigma}\dd {B}^i_t. 
\end{equation}
The force term ${F}^\mathrm{point}_i = {b}({x}_i,\hat{\mu})$ is assumed to depend on the empirical measure 
\[\hat{\mu} = \frac{1}{N}\sum_{j=1}^N \delta_{{x}_j},\]
which encompasses for instance, all binary interaction models. Without the incompressibility force, this system is known as a McKean-Vlasov system and has been studied in great details since the seminal works of McKean and Kac, see the review \cite{chaintron_propagation_2022}. The main idea, which originates from statistical physics, is to consider the limit of the sequence of random empirical measures $\hat{\mu} \equiv \hat{\mu}_N := \frac{1}{N}\sum_{i=1}^N \delta_{{x}_i}$ when $N\to+\infty$. When $\tau=0$ and under some assumptions on ${b}$, it can be shown that this sequence has a deterministic limit $\hat{\mu}_N \to f$ which is the solution of the (nonlinear) Fokker-Planck equation 
\[\partial_t f = - \nabla_{{x}} \cdot ({b}(x,f)f) + \Delta_{{x}}f. \]
The solution $f$ is the probability distribution of the state of a single typical particle and thus models the system at a statistical scale. Many important mathematical results, which are relevant in biology, have been obtained with this method, see for instance \cite{chaintron_propagation_2022} for a review of applications, in particular in collective dynamics. 

Coming back to the Eq.~\eqref{eq:meanfieldincompressible_sm}, a first difficulty comes from the fact that the incompressibility force $\nabla_{{x}_i}\mathcal{T}_c(\hat{\mu})$ does not appear as a simple function of $\hat{\mu}$ due to the gradient in ${x}_i$. To overcome this issue, we recall the notion of first variation of a functional $\mathcal{F}:\mathcal{P}(\Omega)\to\mathbb{R}$ on the space of probability measure $\mathcal{P}(\Omega)$ : this is the function $\frac{\delta \mathcal{F}}{\delta \mu}(\mu) : \Omega \to \mathbb{R}$ defined such that
\[ \frac{\dd}{\dd h} \mathcal{F}(\mu + h\chi)\Big|_{h=0} = \int_{\Omega}\frac{\delta \mathcal{F}}{\delta \mu}(\mu) \dd\chi,\]
for every measure $\chi$ such that $\mu + h\chi \in \mathcal{P}(\Omega)$ for small enough $h$. With this notion, it is possible to apply the chain rule to compute the gradient 
\[\nabla_{{x}_i} \mathcal{T}_c(\hat{\mu}) = N^{-1}\nabla \frac{\delta \mathcal{T}_c}{\delta \mu}(\hat{\mu})({x}_i),\]
where the gradient on the right-hand side is the usual gradient of a real-valued function on $\mathbb{R}^d$. Thus taking the scale $\tau\equiv\tau_N = N\tau_0$, Eq.~\eqref{eq:meanfieldincompressible_sm} can be rewritten, 
\begin{equation}\label{eq:meanfieldfirstvariation}
    \dd {x}_i = {b}({x}_i,\hat{\mu})\dd t - \tau_0 \nabla \frac{\delta \mathcal{T}_c}{\delta \mu}(\hat{\mu})({x}_i)\dd t  + \sqrt{2\sigma_i}\dd {B}^i_t. 
\end{equation}
This latter equation enters the classical framework and leads to the formal mean-field limit 
\begin{equation}\label{eq:FPOT}
    \partial_t f = - \nabla \cdot ({b}({x},f)f) + \tau_0 \nabla \cdot \left( \nabla \frac{\delta \mathcal{T}_c}{\delta \mu}(f) f\right)  + \Delta f.
\end{equation}

At this stage, it remains to identify the first variation of the transport cost $\mathcal{T}_c$. We first note that if $f$ satisfies Eq.~\eqref{eq:FPOT}, it should have a density with respect to the Lebesgue measure. We also assume that the cost is of the form $c({x},{y}) = \ell({x}-{y})$ for a strictly convex function $\ell:\mathbb{R}^d\to[0,+\infty)$. Since the work of Brenier, McCann, Gbangbo, Cafferelli and others, it is well-known that the Monge problem $T\# f = \mathrm{Leb}$ has a unique solution which is of the form 
\[T({x}) = {x} - (\nabla\ell)^{-1}(\nabla\Phi({x})),\]
where $\Phi$ is a Kantorovich potential which satisfies the Monge-Amp\`ere equation \cite{dephilippis_monge_2014}
\[|\det( \nabla T)| = f.\]
Then it can be shown \cite{santambrogio_optimal_2015} that 
\[ \frac{\delta \mathcal{T}_c}{\delta \mu}(f) = \Phi.\]
Consequently, and after a few more computations, Eq.~\eqref{eq:FPOT} can be rewritten as a coupled system of Fokker-Planck-Monge-Amp\`ere equations
\begin{subequations}\label{eq:fpma}
\begin{align}
    &\partial_t f = - \nabla \cdot ({b}({x},f)f) + \tau_0 \nabla \cdot \left( \nabla\Phi f\right)  + \Delta f. \\ 
    &\det\big(\mathrm{I} - \nabla^2\ell^*(\nabla\Phi)\nabla^2\Phi\big) = f,\label{eq:MAgeneral}
\end{align}
\end{subequations}
where $\ell^*$ denotes the Legendre transform of $\ell$ defined by
\[\ell^*({y}) = \sup_{{x}\in\mathbb{R}^d}\{{x}\cdot{y} - \ell({x})\}.\]
For the $L^2$ cost $c({x},{y}) = \frac{1}{2}|{x}-{y}|^2$, it holds that $\ell^*=\ell$ and thus $\nabla^2\ell^*$ is the identity and Eq.~\eqref{eq:MAgeneral} reduces to 
\[\det(\mathrm{I} - \nabla^2\Phi) =  f.\]
With the notable exception of \cite{brenier_geometric_2004} which discusses connection with the Euler equation, the general equation \eqref{eq:fpma} is not common in the literature and, to the best of our knowledge, it is new in a mathematical biology framework. Its mathematical analysis and numerical treatment remain largely open, as well as the study of different scaling limits and the influence of the cost function. This may represent a novel rigorous approach to derive continuum PDE models from systems of particles with volume interactions. Connections with continuum model of crowd motion \cite{leclerc_lagrangian_2020,natale_gradient_2023} are also natural although the authors of these works follow a different path than mean-field theory.

\newpage

\begin{figure}
    \centering
    \includegraphics[width=0.8\textwidth]{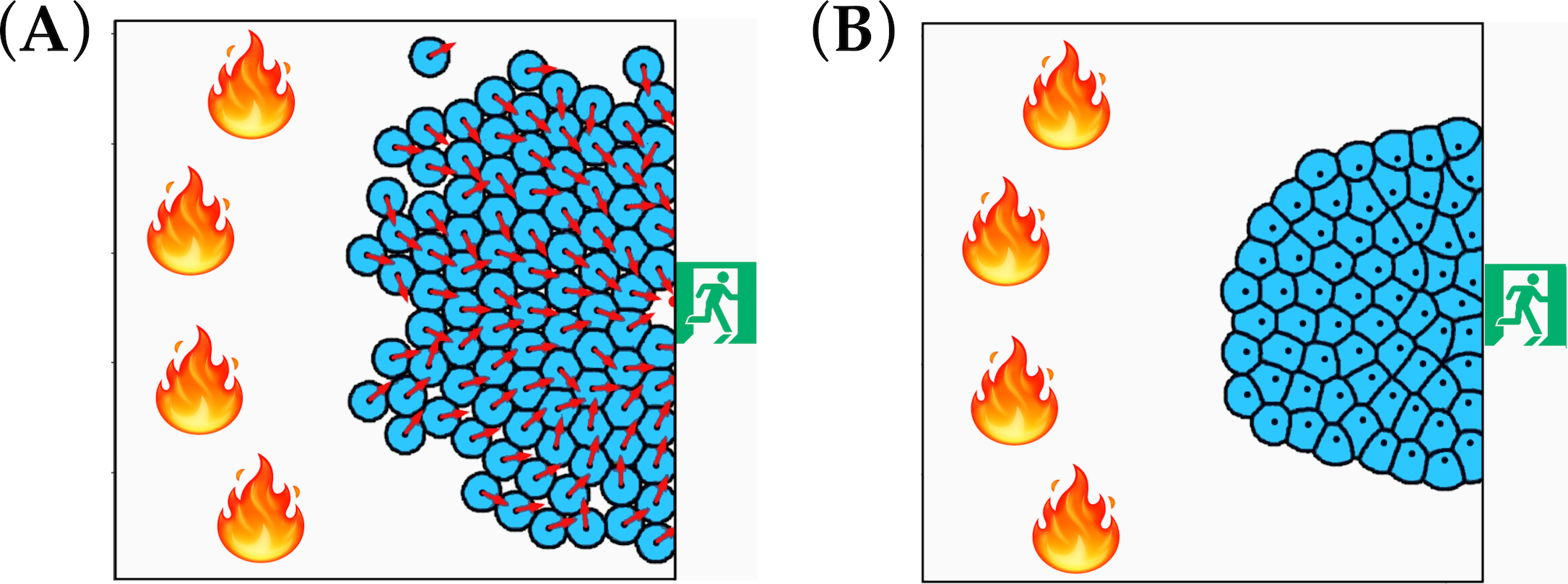}
    \caption{\textbf{Crowd simulations.} \textbf{(A)} $\alpha=8$. Formation of a stable arch around the exit point when deformations are not allowed. \textbf{(B)} $\alpha=0.5$. The crowd is able to leave the room when deformations are allowed. See also SM Videos~\ref{video:crowd_a05},\ref{video:crowd_a8}.}
    \label{fig:evacuation}
\end{figure}

\begin{figure}
    \centering
    \includegraphics[width=0.75\textwidth]{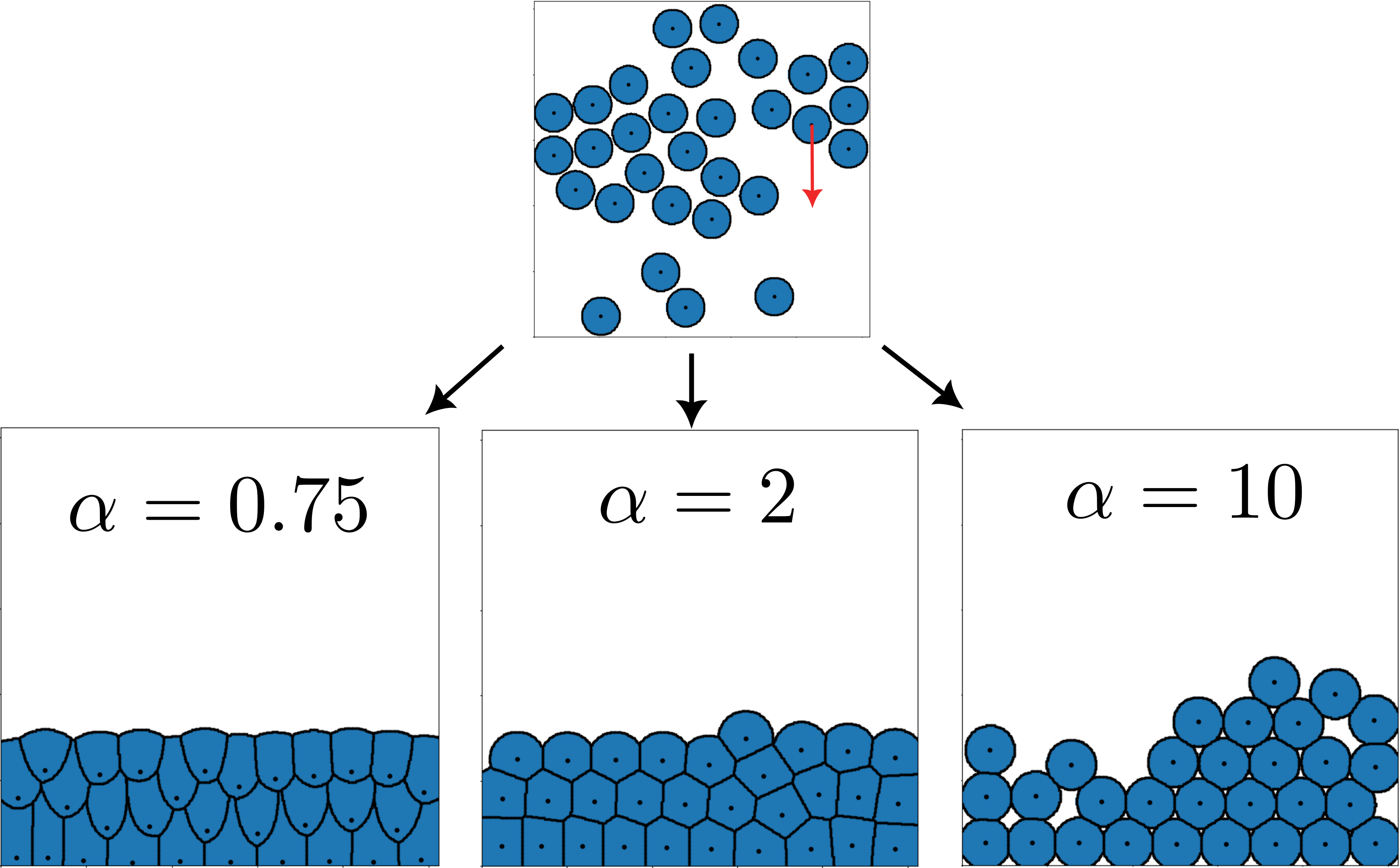}
    \caption{\textbf{Falling soft spheres.} Initial configuration (top) and equilibrium configuration (bottom) of a system of falling soft spheres for three values of the deformation parameter $\alpha$. See also SM Videos~\ref{video:fs_a075},\ref{video:fs_a2},\ref{video:fs_a10}.}
    \label{fig:fallingspheres}
\end{figure}

\begin{figure}
    \centering
    \includegraphics[width=0.9\textwidth]{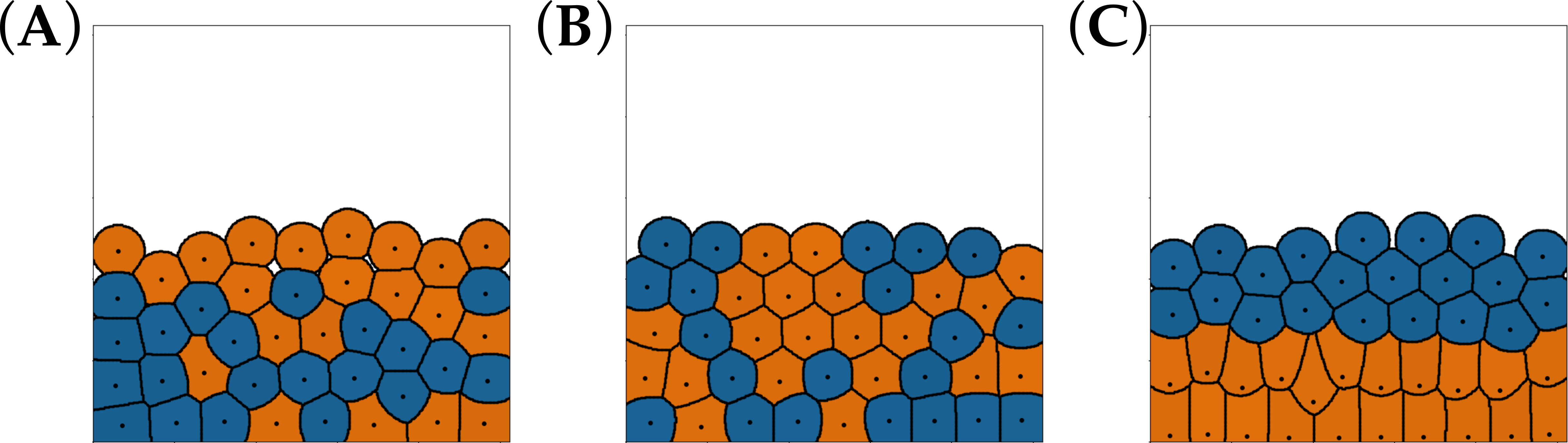}    
    \caption{\textbf{Final configuration of a mixed system of 30 falling spheres with different weights and softness.} The 15 blue spheres are harder ($\alpha=2$) and the orange spheres are softer ($\alpha=1$). The gradient descent step of the blue spheres is fixed to $\tau_b=3$ and both the orange and blue spheres are subject to the same downward force with magnitude $F_0=0.4$. The gradient descent step of the orange particles $\tau_o$ is analogous to the inverse of a mass. (a) Light orange spheres ($\tau_o=8$) (b) Same weight ($\tau_o=2$) (c) Heavier orange spheres ($\tau_o=1$). See also SM Videos~\ref{video:fs_tau8},\ref{video:fs_tau3},\ref{video:fs_tau1}.}
    \label{fig:fallingspheres_mixed}
\end{figure}

\begin{figure}
    \centering
    \includegraphics[width=0.7\textwidth]{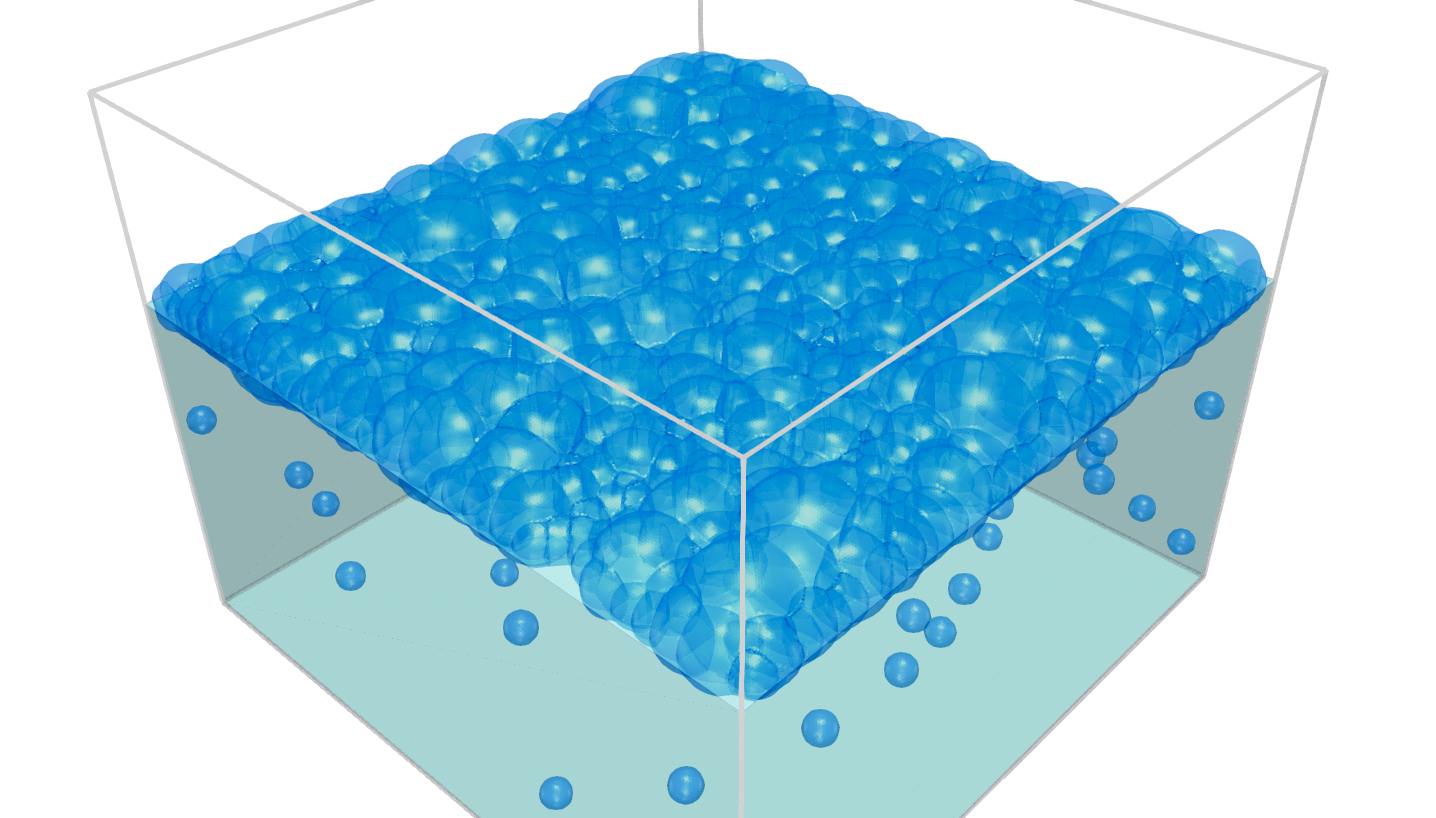}
    \caption{\textbf{Bubble simulations.} See also SM Video~\ref{video:bubbles}.}
    \label{fig:bubbles}
\end{figure}

\newpage

\subsection*{Supplementary figures}

\begin{figure}
    \centering
    \includegraphics[width=0.85\linewidth]{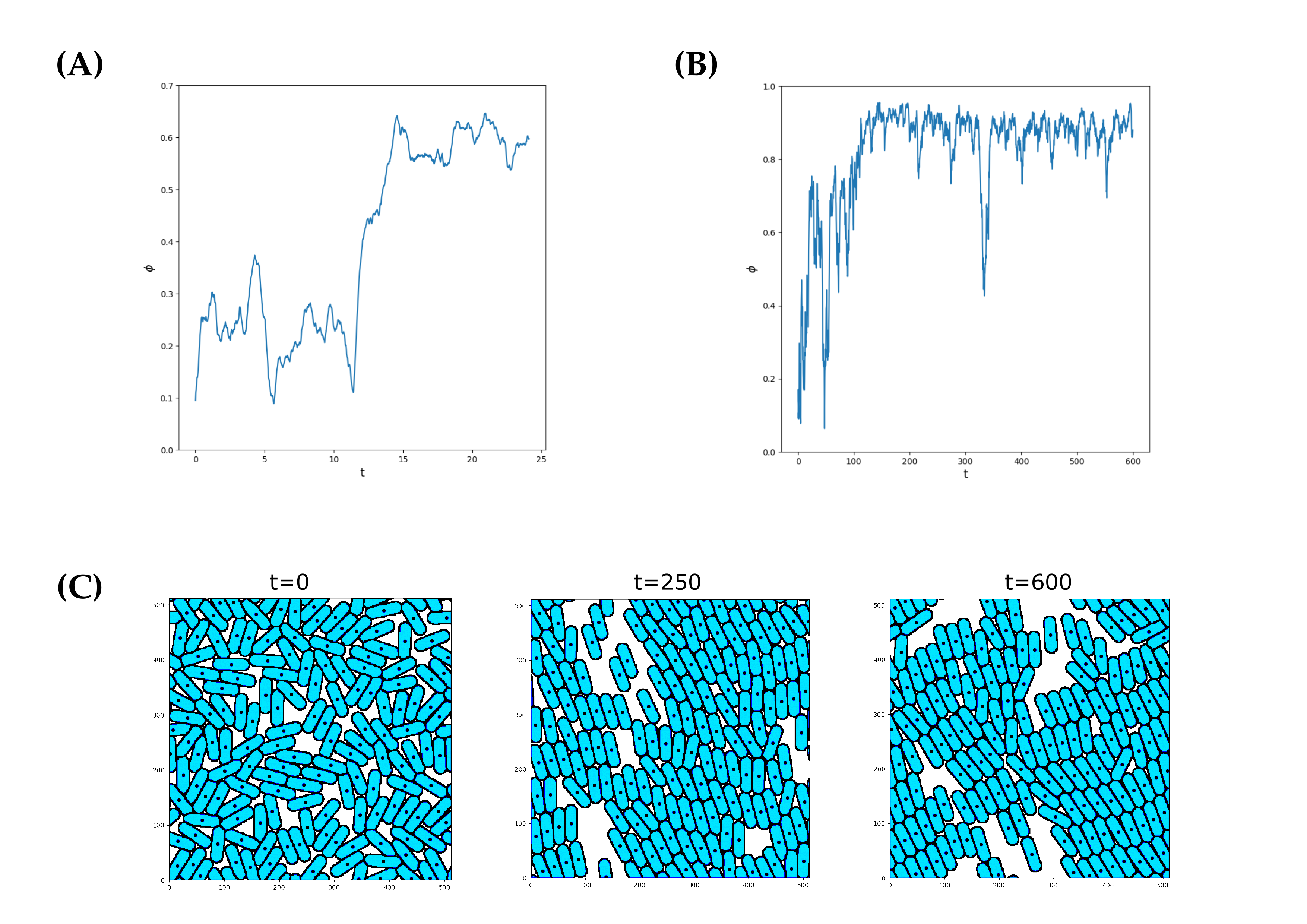}
    \caption{\textbf{Nematic order parameter over time in the simulation of rod-shape particles.} \textbf{(A)} Related to Fig.~\ref{fig:shapes}B
    \label{fig:nematicorder} \textbf{(B)} Nematic order parameter over an extended simulation time $(T=600)$ for a smaller but denser system ($N=150$, $V=0.8$, $\alpha=8$). The other parameters are the same as Fig.~\ref{fig:shapes}B. \textbf{(C)} Snapshots of the simulation (B) at three time points showing the persistence of orientational order.}
\end{figure}

\begin{figure}
    \centering
    \includegraphics[width=0.92\linewidth]{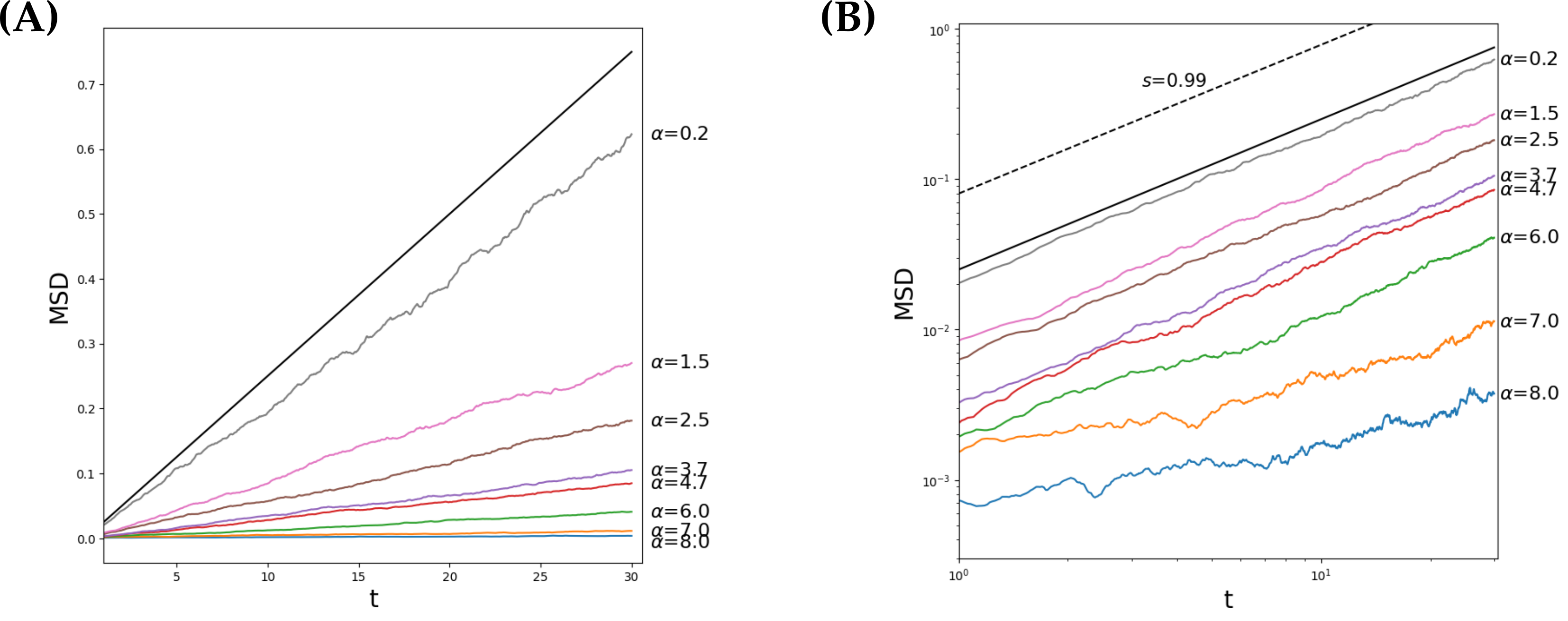}
    \caption{\textbf{MSD for Active Brownian Particles with various deformability parameters.} \textbf{(A)} Linear scale until time $t=30$ for eight deformability parameters $\alpha$ and $c_0=0.5$. The solid black line has the theoretical slope for non-interacting ABM. \textbf{(B)} Log-log scale plot for $t\in(1,30)$. The average anomalous diffusion coefficient computed for the fluid-like systems in the range $\alpha\leq6$ is $s=0.99$ (dashed line).}
    \label{fig:msdloglog}
\end{figure}

\begin{figure}
    \centering
    \includegraphics[width=0.85\linewidth]{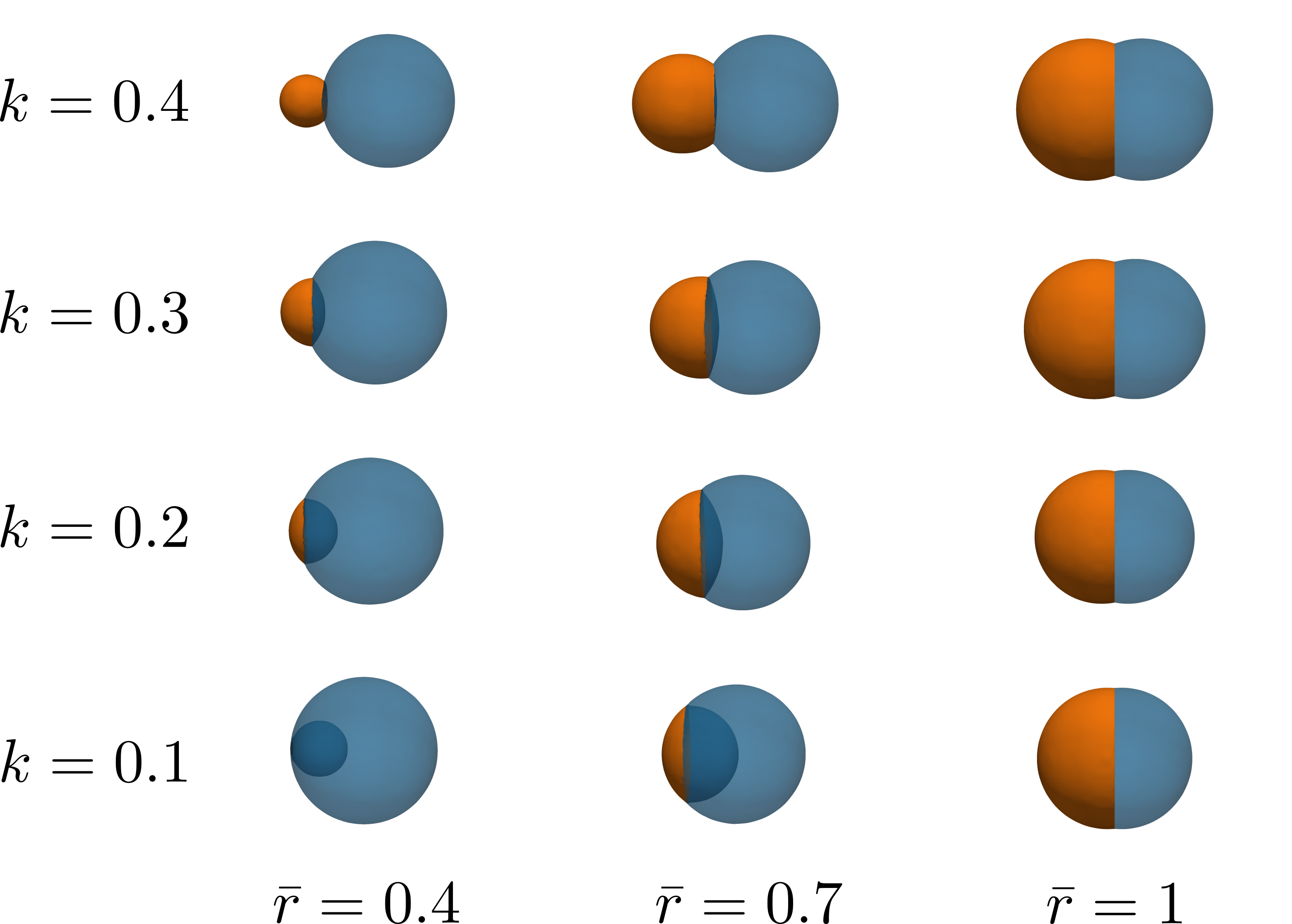}
    \caption{\textbf{Equilibrium configurations for two 3D cells with different volumes.} Simulation of two 3D Laguerre cells with different volume $v_o$ and $v_b$ (resp. orange and blue) until an equilibrium configuration of the model \eqref{eq:interfaceforce} is reached. The blue cell is displayed with lower opacity. We define the ratios ${k} = \frac{\eta_{ob}}{2\gamma_o}$ and $\bar{r}=(v_o/v_b)^{1/3}$. The fixed parameters are $\gamma_o=\gamma_b=7$, $\gamma_{ob}=1$ and we set $\gamma_{oo}=\gamma_{bb}=\eta_{oo}=\eta_{bb}=0$ since there is only one orange/blue interface.}
    \label{fig:compaction}
\end{figure}


\newpage

\subsection*{Supplementary tables}

\begin{table}[ht]
\tiny
    \centering
        \caption{\textbf{Modeling and numerical parameters.} This table gathers all the detailed modeling and numerical parameters used in the experiments. When the parameters differ between blue and orange particles, they are identified respectively by the subscripts $b$ and $o$. The parameters $\lambda$ and $b$ refer to the cost normalization procedure with $\lambda=1$ if not specified. When a single value of $V$ is given, it corresponds to the total volume of the $N$ particles which are all assumed to have the same volume $V/N$. The space domain $\Omega$ is discretized with $M$ voxels in dimension $d$. The value $\Delta t$ corresponds to the time discretization step in the Euler method. The vector ${e}_z$ denotes the unit vector pointing upward. The operator $\mathsf{P}_{{n}^\perp}$ denotes the projection on the orthogonal of a unit vector ${n}$ and ${B}^i_t$ is a Brownian motion. The point ${x}_\mathrm{exit} = (1,0.5)^\mathrm{T}$ denotes the exit point in Fig.~\ref{fig:evacuation}. The notation $x=(x_1,x_2,x_3)$ indicates the values for the respective figures indexed by $(a,b,c)$ or from left to right. The notation $x\in(x_1,x_2)$ indicates that the value of $x$ is sampled uniformly (randomly if not indicated) within the interval $(x_1,x_2)$ for each particle.}

\begin{tabular}{p{0.06\textwidth}||p{0.08\textwidth}|p{0.08\textwidth}|p{0.04\textwidth}|p{0.04\textwidth}|p{0.17\textwidth}||p{0.08\textwidth}|p{0.08\textwidth}||p{0.04\textwidth}|p{0.04\textwidth}}
\\

         Experiment & cost  & ${F}^\mathrm{point}$ & ${F}^\mathrm{surf}$  & $\tau$ & other  & $N$  & $V$  & $\Delta t$  & $M^{1/d}$ \\
         \hline\hline
         
         Fig.~\ref{fig:benchmark}a & {$L^2$} \newline {$\lambda=0.25$} & $-0.5{e}_z$  & 0  & 3  & NR & 50 to 2,000 & 0.5 & 0.005 & 512 \\
         \hline
         
         Fig.~\ref{fig:benchmark}b& {$L^2$} \newline {$\lambda=0.1$}  & 0 & 0 & 10 & growth rate: $3\gamma v_0$ \newline $\gamma \in (0.5,2)$ random \newline division rate: 3 & 1 to 50,000 & {target volume:} \newline $v_0=\frac{4}{3}\pi0.15^3$  & 0.002 & 400 \\
         \hline
         
         Fig.~\ref{fig:benchmark}c& {\eqref{eq:ellipsoidcost}} \newline $\lambda\in(3,4)$, $b=1$  & $0.3{n}_i$ & 0  & 42  & aspect ratio: 1 to $4$  & 10 to 10,000  & {$V=0.8$} \newline random with ratio 1/5  & 0.005  & 64 to 512 \\
         \hline
         
         Fig.~\ref{fig:shapes}b& \eqref{eq:costshape}-\eqref{eq:spherocylindercost}\newline $\alpha=3.5$ & $0.3{n}_i$ & 0 & 0.14  & aspect ratio: 3  & 300  & $0.65$  & 0.002  & 512 \\
         \hline
         
         Fig.~\ref{fig:abp}& \eqref{eq:powercost} $\alpha\in(0.2,8)$ \newline {$\lambda=0.1$} & {$c_0{n}_i$} \newline {$\dot{{n}}_i = \sqrt{2\sigma} \mathsf{P}_{{n}_i^\perp}\circ \dot{{B}}^i_t$} & 0 & 3 & {$\sigma=20$} \newline {$c_0\in(0.02,2)$} & 250 & 0.94 & 0.002 & 512 \\
         \hline

         Fig.~\ref{fig:shapes}c& \eqref{eq:costpotentialchemotaxis} & 0 & 0 & 1 & {$\beta=0.02$, \eqref{eq:bias}} \newline $u$ Gaussian with \newline variance $0.2^2$ & 10 & 0.1 & 0.01 & 512 \\
         \hline

         Fig.~\ref{fig:3Dsorting}a& \eqref{eq:interfaceforce} & 0  & {\eqref{eq:interfaceforce}}  & 0  & $\overline{\eta}=3$, $k_2=k_{12}=0.4$, $\gamma_2=10$ \newline $\overline{\gamma} = 1/3$, $\overline{k} = (0.06,0.21,0.67,1,2)$ \newline  $\overline{\gamma} = 1$, $\overline{k} = (0.17,0.67)$ & {$N_o = 60$} \newline {$N_b=60$} & 0.27  & 0.001 & 256 \\
         \hline

         Fig.~\ref{fig:3Dsorting}b& \eqref{eq:interfaceforce} & 0  & {\eqref{eq:interfaceforce}}  & 0  & $\overline{\eta}=0.3$, $k_2=k_{12}=0.4$, $\gamma_2=10$ \newline $\overline{k} = 0.5$, $\overline{\gamma} = (1,1.3,3.1,13.3)$ \newline  $\overline{k} = 1$, $\overline{\gamma} = (1,1.5,6.7)$ \newline $\overline{k}=2$, $\overline{\gamma}=(1.25,3.3)$ & {$N_o = 60$} \newline {$N_b=60$} & 0.27  & 0.001 & 256 \\
         \hline

         Fig.~\ref{fig:3Dsorting}c& \eqref{eq:interfaceforce} & 0  & {\eqref{eq:interfaceforce}}  & 0  & {$\overline{\eta}=3$, $\overline{\gamma}=2$, $\overline{k}\overline{\gamma}\overline{\eta}=0.8$} & {$N_o = 64$} \newline {$N_b=64$} & 0.29  & 0.0003 & 256

    \end{tabular}
    \label{tab:parameters}
\end{table}

\begin{table}[ht]
\tiny
    \centering
        \caption{\textbf{Modeling and numerical parameters for the Supplementary experiments}}

    \begin{tabular}{p{0.06\textwidth}||p{0.08\textwidth}|p{0.08\textwidth}|p{0.04\textwidth}|p{0.04\textwidth}|p{0.17\textwidth}||p{0.08\textwidth}|p{0.08\textwidth}||p{0.04\textwidth}|p{0.04\textwidth}}
    		\\

         Experiment & cost  & ${F}^\mathrm{point}$ & ${F}^\mathrm{surf}$  & $\tau$ & other  & $N$  & $V$  & $\Delta t$  & $M^{1/d}$ \\
         \hline\hline

         Fig.~\ref{fig:evacuation}A,B& \eqref{eq:powercost}, $\alpha=(0.5,8)$ \newline $\lambda=0.1$ & $0.4{n}_i$ \newline ${n}_i = ({x}_i - {x}_\mathrm{exit})/|{x}_i - {x}_\mathrm{exit}|$ & 0  & 3 & $\sigma=0.2$  & 111 & 0.42 & 0.001 & 512 \\
         \hline
         
         Fig.~\ref{fig:fallingspheres}& \eqref{eq:powercost} $\alpha=(0.75,2,10)$ \newline $\lambda=0.125$ & $-0.4{e}_z$  & 0 & 1.5  & NR & 30 & 0.3 & 0.001 & 512 \\
         \hline
         
         Fig.~\ref{fig:fallingspheres_mixed}A,B,C& \eqref{eq:powercost}, $\alpha_b=2$, $\alpha_o=1$ \newline $\lambda=0.125$ & $-0.4{e}_z$ & 0 & $\tau_b=3$ \newline $\tau_o=(1,3,8)$  & NR & $N_b=N_o=21$ & 0.5 & 0.001 & 512 
         
    \end{tabular}
    \label{tab:parameters_supplementary}
\end{table}


\clearpage 
%
%

\subsection*{List of Supplementary Videos}

The following supplementary videos can be found on Figshare 
\begin{center}
\url{https://doi.org/10.6084/m9.figshare.25240669}
\end{center}

or on the documentation website alongside with the related Python scripts (and additional simulations not shown in the present article). In all the videos, the volume of the domain is normalized to 1 but the ticks on the axes are labeled between $1$ and $M$ (the number of discrete voxels per dimension). 

\begin{center}
    \url{https://iceshot.readthedocs.io/}
\end{center}

\begin{video}[Incompressibility]\label{video:incompressibility} Toy simulation of the gradient descent dynamics \eqref{eq:incompressibilityforce} in a four-cell system. The negative gradient of the cost function (called incompressibility force) is displayed in red.
\end{video}

\begin{video}[3D hourglass]\label{video:fallingstuff} Falling soft spheres in a hourglass domain with a spherical obstacle. Particles are progressively added in the funnel. The color of a particle only indicates the time when it has been added. See also Fig.~\ref{fig:benchmark}A. 
\end{video}

\begin{video}[3D tissue growth]\label{video:tissue_growth_3D} Growing cell aggregate in 3D following a basic somatic cell division process. See also Fig.~\ref{fig:benchmark}B. 
\end{video}

\begin{video}[Benchmark: deformable ellipsoids]\label{video:benchmark} Benchmark experiment introduced in the main text for a system of $N=1000$ self-propelled deformable ellipsoids with a discretization grid of size $M=512^3$. See also Fig.~\ref{fig:benchmark}C. 
\end{video}

\begin{video}[Rod-shape particles]\label{video:rodshape} Related to Fig.~\ref{fig:shapes}B: simulation of a system of rod-like particles with directional changes induced by deformation and bending effects. Long-range order spontaneously emerges after some time. 
\end{video}

\begin{video}[Active Brownian Particles (1/3)]\label{video:abp_a05} Related to Fig.~\ref{fig:abp}B: simulation of a system of active Brownian particles with deformability parameter $\alpha=0.5$. The mean-square displacement of the particles is smaller but of the same order of magnitude than the theoretical value for independent point particles. 
\end{video}

\begin{video}[Active Brownian Particles (2/3)]\label{video:abp_a2} Related to Fig.~\ref{fig:abp}B: simulation of a system of active Brownian particles with deformability parameter $\alpha=2$. As the deformability decreases, the mean-square displacement is drastically reduced.
\end{video}

\begin{video}[Active Brownian Particles (3/3)]\label{video:abp_a10} elated to Fig.~\ref{fig:abp}B: simulation of a system of active Brownian particles with deformability parameter $\alpha=10$. The particles behave as hard-spheres and self-organize into an optimal sphere packing configuration with almost no movement. 
\end{video}

\begin{video}[Chemotaxis motion (1/2)]\label{video:chemo_long} Related to Fig.~\ref{fig:shapes}C: simulation of a system of 10 particles where the movement is induced by the shape deformation in response to a chemo-attractant gradient. With a particular choice of cost perturbation, particle migrate by adopting elongated shapes in the direction of increasing chemo-attractant gradient. 
\end{video}

\begin{video}[Chemotaxis motion (2/2)]\label{video:chemo_fan}Related to Fig.~\ref{fig:shapes}C: simulation of a system of 10 particles where the movement is induced by the shape deformation in response to a chemo-attractant gradient. With a particular choice of cost perturbation, particle migrate by adopting fan-like shapes orthogonal to the direction of increasing chemo-attractant gradient. 
\end{video}

\begin{video}[3D Cell sorting (1/3)]\label{video:st_separation} Related to Fig.~\ref{fig:3Dsorting}A: simulation of a homogeneously mixed two-population cell aggregate with $N=128$ and $\overline{\eta}=3$, $\overline{\gamma} = 1$ and $\overline{k} = 1$ leading to the separation of the two populations. 
\end{video}

\begin{video}[3D Cell sorting (2/3)]\label{video:st_checkerboard} Related to Fig.~\ref{fig:3Dsorting}B: simulation of a homogeneously mixed two-population cell aggregate with $N=128$ and $\overline{\eta}=0.3$, $\overline{\gamma} = 1$ and $\overline{k} = 1$ leading to a checkerboard pattern state.
\end{video}

\begin{video}[3D Cell sorting (3/3)]\label{video:st_internalization} Related to Fig.~\ref{fig:3Dsorting}A: simulation of a homogeneously mixed two-population cell aggregate with $N=128$ and $\overline{\eta}=3$, $\overline{\gamma} = 2$ and $\overline{\eta}\overline{\gamma}\overline{k} = 0.8$ leading to the internalization of the hardest cells (region A).
\end{video}

\begin{video}[3D Engulfment]\label{video:st_engulfment} Related to Fig.~\ref{fig:3Dsorting}C: simulation of a homogeneously mixed two-population cell aggregate with $N=128$ and $\overline{\eta}=3$, $\overline{\gamma} = 2$ and $\overline{\eta}\overline{\gamma}\overline{k} = 0.8$ leading to the engulfment of the hardest cells (region A) even with an initially totally segregated state. 
\end{video}

\begin{video}[Crowd motion (1/2)]\label{video:crowd_a05} Related to Fig.~\ref{fig:evacuation}: simulation of crowd motion towards a single exit point with deformability parameter $\alpha=0.5$. All the particles manage to escape at the price of large deformations. 
\end{video}

\begin{video}[Crowd motion (2/2)]\label{video:crowd_a8} Related to Fig.~\ref{fig:evacuation}: simulation of crowd motion towards a single exit point with deformability parameter $\alpha=8$. The particles behave as hard-spheres and end up in a stable congested arch state. 
\end{video}

\begin{video}[Falling spheres (1/6)]\label{video:fs_a075} Related to Fig.~\ref{fig:fallingspheres}: falling spheres with the deformation parameter $\alpha=0.75$. Initially round shape cells adopt a columnar shape due to the external force and a high tolerance to deformation.
\end{video}

\begin{video}[Falling spheres (2/6)]\label{video:fs_a2} Related to Fig.~\ref{fig:fallingspheres}: falling spheres with the deformation parameter $\alpha=2$. The initially round particles are not so prone to deformation: they thus keep a convex shape and organize into a Voronoi-like configuration. 
\end{video}

\begin{video}[Falling spheres (3/6)]\label{video:fs_a10} Related to Fig.~\ref{fig:fallingspheres}: falling spheres with the deformation parameter $\alpha=10$. The particles behave as hard-spheres. 
\end{video}

\begin{video}[Falling spheres (4/6)]\label{video:fs_tau8} Related to Fig.~\ref{fig:fallingspheres_mixed}: falling spheres where the magnitude of the incompressibility force of the orange and blue spheres are respectively $\tau_o=8$ and $\tau_b=3$.
\end{video}

\begin{video}[Falling spheres (5/6)]\label{video:fs_tau3} Related to Fig.~\ref{fig:fallingspheres_mixed}: falling spheres where the magnitude of the incompressibility force of the orange and blue spheres are respectively $\tau_o=3$ and $\tau_b=3$.
\end{video}

\begin{video}[Falling spheres (6/6)]\label{video:fs_tau1}Related to Fig.~\ref{fig:fallingspheres_mixed}: falling spheres where the magnitude of the incompressibility force of the orange and blue spheres are respectively $\tau_o=1$ and $\tau_b=3$.
\end{video}

\begin{video}[Bubbles]\label{video:bubbles}Related to Fig.~\ref{fig:bubbles}: Simple simulation of a system of bubbles with random creation, explosion and fusion events at the interface between a fluid and the air.
\end{video}



\end{document}